\documentclass[twocolumn,pre,superscriptaddress,nofootinbib]{revtex4}

\usepackage{graphicx}
\usepackage{color}
\usepackage{natbib}
\usepackage[colorlinks=true,allcolors=blue]{hyperref}
\usepackage[T1]{fontenc}

\begin{document}

\title{Directedeness, correlations, and daily cycles in springbok motion:
from data over stochastic models to movement prediction}

\author{Philipp G. Meyer}
\affiliation{University of Potsdam, Institute of Physics \& Astronomy,
14476 Potsdam-Golm, Germany}
\author{Andrey G. Cherstvy}
\affiliation{University of Potsdam, Institute of Physics \& Astronomy,
14476 Potsdam-Golm, Germany}
\author{Henrik Seckler}
\affiliation{University of Potsdam, Institute of Physics \& Astronomy,
14476 Potsdam-Golm, Germany}
\author{Robert Hering}
\affiliation{University of Potsdam, Plant Ecology and Nature Conservation,
14469 Potsdam, Germany}
\author{Niels Blaum}
\affiliation{University of Potsdam, Plant Ecology and Nature Conservation,
14469 Potsdam, Germany}
\author{Florian Jeltsch}
\email{jeltsch@uni-potsdam.de}
\affiliation{University of Potsdam, Plant Ecology and Nature Conservation,
14469 Potsdam, Germany}
\affiliation{Berlin-Brandenburg Institute of Advanced Biodiversity Research,
14195 Berlin, Germany}
\author{Ralf Metzler}
\email{rmetzler@uni-potsdam.de}
\affiliation{University of Potsdam, Institute of Physics \& Astronomy,
14476 Potsdam-Golm, Germany}
\affiliation{Asia Pacific Center for Theoretical Physics, Pohang 37673,
Republic of Korea}

\date{\today}

\begin{abstract}
How predictable is the next move of an animal? Specifically, which factors
govern the short- and long-term motion patterns and the overall dynamics
of landbound, plant-eating animals and ruminants in particular? To answer
this question, we here study the movement dynamics of springbok antelopes
\textit{Antidorcas marsupialis}. We propose complementary statistical
analysis techniques combined with machine learning approaches to analyze,
across multiple time scales, the springbok motion recorded in long-term
GPS-tracking of collared springboks at a private wildlife reserve in Namibia.
As a new result, we are able to predict the springbok movement within the
next hour with a certainty of about 20\%. The remaining 80\% are stochastic
in nature and are induced by unaccounted factors in the modeling algorithm
and by individual behavioral features of springboks. We find that directedness
of motion contributes approximately 17\% to this predicted fraction. We find
that the measure for directedeness is strongly dependent on the daily cycle.
The previously known daily affinity of springboks to their water points, as
predicted from our machine learning algorithm, overall accounts for only 3\%
of this predicted deterministic component of springbok motion. Moreover, the
resting points are found to affect the motion of springboks at least as much
as the formally studied effects of water points. The generality of these
statements for the motion patterns and their underlying behavioral reasons
for other ruminants can be examined on the basis of our statistical analysis
tools.
\end{abstract}

\maketitle

\section{Introduction}

Ronald Ross, who received the 1902 Nobel Prize for Physiology or Medicine for
his discovery on the transmission of malaria by mosquitoes, formulated the
fundamental problem to understand the spatiotemporal spreading of infected
mosquitoes from a breeding pool \cite{ross}: "Suppose that a mosquito is born
at a given point, and that during its life it wanders about, to and fro, to
left or to right, where it wills, in search of food or of mating, over a
country which is uniformly attractive and favorable to it. After a time it
will die. What are the probabilities that its dead body will be found at a
given distance from its birthplace?" To solve this problem, Ross contacted
the statistician Karl Pearson, who is credited for creating the concept of
the random walk. In an exchange of Letters in Nature with John William Strutt,
Lord Rayleigh, it became apparent that after sufficiently many steps the
position of the random walker converges to a Gaussian random variable
\cite{pearson,rayleigh,pearson1,benh08walks,hughes}. In fact, the random walk
is close to Einstein's and Smoluchowski's formulations of the mathematical
theory of diffusion \cite{einstein,smoluchowski}, and generalizations of
this simple concept have been extensively used in the modeling of how animals
search for resources.

%A milestone towards modern movement ecology were animal counts and the tracking
%of their seasonal migration patterns by aerial observation, in Bernhard and
%Michael Grzimek's campaign in the Serengeti (Tanzania, Africa). From their
%data the Grzimeks recommended the best outline for the newly established
%Serengeti National Park. The Grzimeks' documentary "Serengeti shall not
%die" on their work won an Academy Award in 1959 and their book became an
%international bestseller \cite{grzimek}.

A milestone towards modern movement ecology were animal counts and the tracking
of their seasonal migration patterns by aerial observation in the late 1950s,
informing authorities on the best layout for the newly established Serengeti
National Park in Tanzania, Africa.
Today, several well-developed methods to record the movement of animals are
routinely employed \cite{benson}. We mention GPS tracking of transmitting tags
\cite{gps,gps1,jelt22} and automated radio tracking, particularly, the
high-throughput ATLAS (Advanced Tracking and Localization of Animals in
real-life Systems) \cite{atlas,atlas1}. The observations garnered by such
methods are at the core of movement ecology, an emerging unifying paradigm
for understanding the various mechanisms underlying animal behavior and the
consequences for key ecological and evolutionary processes \cite{nath08para,
Nathan22}. Thus movement ecology builds on early theoretical ideas of dispersal
in populations \cite{pear06,skel51} and the connection with epidemic
spreading \cite{brow12epid}. Classes of interest in movement ecology based on
individual tracking of organisms include, i.a., mammals \cite{jelt22,
tuck18}, birds \cite{assa22prx,vilk,atlas1,jelt16storks}, bats
\cite{roeleke22}, or marine predators \cite{sims08,sims11}. A key question is
to predict the "next move" \cite{gura16tortous} of an animal with some
statistical certainty.

Unveiling a broader picture in the movement ecology of even a single species is
hampered by various factors, including effects of seasonal changes of animal
behavior and the environment, migration, reproduction cycles and breeding,
intra-species collective phenomena and group size, interactions with foreign
species, variations of behaviors and physiology between individuals,
heterogeneity of the environment and resources, home-ranging and confinement,
among a variety of other factors. The original expectations that relatively
simple stochastic models could capture essential features of movement ecology
therefore have remained elusive, and most models are species-specific. While
a universal modeling framework may be unattainable, a central longer-term
question is whether we can identify universal dynamic modules, which occur
for broad ranges of species.

Several stochastic models have been used to describe the movement patterns
of animals, starting with the normal random walk (Pearson walk) mentioned
above for the description of mosquito spreading and applications to crabs
\cite{pear06} and muskrats \cite{skel51}. While a random walk is a
Markovian process and the direction of each jump independent of the previous
jump, a direct generalization is represented by correlated random walks with
a finite correlation time (typically modeled with Ornstein-Uhlenbeck noise or
by coupling to a diffusive rotational motion of the direction of motion)
\cite{fuer20,kare83,visw05corr,benh08walks,romanczuk2012,
gura17corrvelo,elisabeth}. Beyond the correlation time, such processes are
normal-diffusive with an effective diffusivity \cite{romanczuk2012}. Once
the correlations are taken to be long-range, such as in fractional
Brownian motion with positive, power-law correlations of the driving
Gaussian noise \cite{mandelbrot}, the resulting motion is superdiffusive
with a mean squared displacement (MSD) of the form $\langle\mathrm{r}^2(t)
\rangle\simeq t^{\alpha}$ and $1<\alpha\le2$ \cite{report,metz14,soko15}.
Crossovers from superdiffusive power-law forms of the MSD to normal
diffusion or another power-law can be achieved by different forms of
tempering of the driving noise \cite{daniel}. In the standard formulation
of these models the motion is unbounded. Finite domains such as home ranges
or confinement by geographic boundaries or fencing can then be included by
appropriate boundary conditions or introduction of a confining potential.
In such cases the motion will eventually reach a non-equilibrium steady
state (NESS) characterized by a stationary probability density function
(PDF).\footnote{An exception is given by persistent fractional Brownian
motion when the confining potential is too weak \cite{tobiasg}.} A NESS
can also be reached by so-called random resetting strategies, in which an
a priori unbounded stochastic process such as Brownian motion is "reset"
back to its origin (or to another point with a given probability
\cite{marcus,shlomi}), typically following a Poisson statistic of reset
times \cite{satya,evansreview}. Resetting has been generalized to a large
number of stochastic processes, we only mention some representative
examples \cite{activereset,reset1,reset2,trifce,trifce1,andrey,andrey1,
chechkin,chechkin1}.

In the context of random search for sparse food a long debate has focused
on "optimal search" \cite{olivierreview,searchreview}. Intermittent search
strategies \cite{olivier,olivier1,gandhi,michael,vladinter} combine local
search, typically Brownian motion, with a persistent process such as
ballistic motion. The role of the latter is to decorrelate the searcher by
relocating it to a remote area, that likely has not been visited before. A
second strategy to reduce oversampling in one and two dimensions are L{\'e}vy
search processes, in which relocation lengths are power-law distributed, such
that hierarchical clusters are searched \cite{mandelbrotbook,shlekla}. This
reduces the search time \cite{shlekla,michael,michaelprl}. The L{\'e}vy flight
foraging hypothesis \cite{gandhi} led to a large number of studies identifying
L{\'e}vy patterns in animal motion \cite{visw96,visw99,stan07,sims12,Gurarie09,
gura22,boye06,boye14,sims08,hump10tuna,visw10,sims11,jage11,foca17,kare83,
hirt22,mura20bugs} and in human motion patterns \cite{raic14human,brockmann,
talkoren,havlincovid}. While in some cases the L{\'e}vy model has been
questioned \cite{pyke15,stan07,trav07,edwa12,paly14levy}, it focused the
interest of the statistical physics community on movement ecology.

Obviously, while simple random search models may provide essential insight
into observed motion patterns, they cannot capture the full complexity
displayed by higher animals and thus represent a starting point for further
analysis. To develop better models in movement ecology, various other effects
need to be considered, e.g., spatial memory \cite{faga13},
task-optimized navigation strategies \cite{panzachhi2015reindeer}, food
abundance \cite{CHEN202215}, group dynamics \cite{roeleke22}, information
exchange between sub-populations in a community \cite{jelt20rev-commun,roeleke22},
heterogeneity of the terrain, habitat-connectivity \cite{panzachhi2015reindeer},
or selection \cite{potts2020diffssa}. We here propose dedicated statistical
analysis techniques in combination with machine-learning (ML) approaches, to study
to which degree these tools can be used towards predicting the movement
patterns of ruminants depending, inter alia, on daily and annual cycles, on
resource distribution in the home range, on area restriction due to confinement,
and on additional features, that can only be described in a statistical way. We
validate these methods to analyze long-term GPS-tracking data of springboks
in a private wildlife reserve in Namibia on various time-scales. We show that
our approach helps us to understand the limitations, but also some important
features of springbok motion
which can be generalized to the description of motion patterns of other ruminant
species inhabiting semi-arid landscapes. Moreover, the concepts developed here
are promising to be applied on a more generic level to analyze and predict
movement patterns of higher animals.

The structure of the paper is as follows. We introduce the data set for our
study in Sec.~\ref{sec_data} and characterize it by applying stochastic
modeling in Sec.~\ref{sec_tsd}. In Sec.~\ref{sec_param} we focus on two main
questions, the importance of water point positions for the overall dynamics
of springbok movement and how the directedness \cite{benh04-tortuos-straight}
of gazelle motion and their activity depend on the hour of the day. Finally, in
Sec.~\ref{sec_pred} we discuss the forecasting power of the ML-based models
in dependence on input features such as season, water point distance, and
vegetation levels. We conclude our investigation in Sec.~\ref{conc}.

\section{Data acquisition and visualization}
\label{sec_data}

The movement patterns of medium-sized ruminants such as antelopes, gazelles,
and springboks have been intensely investigated \cite{walt68,Bigalke1970,
bigalke1972,davi78,skin96}. While earlier studies relied on extensive direct
observations such as aerial monitoring or radio tracing by handheld antennae,
automated
contemporary tracking methods allow scientists to garner high-resolution,
long-time tracking data. An extensive range of behavioral details has been
revealed, including activity rhythms, seasonal influence, water needs, gaits,
feeding habits and preferences including, social habits, reproduction cycles,
lambing peaks, herd composition, age-related changes, sex-specific behavior,
body-weight distribution, etc. \cite{Bigalke1970,bigalke1972,davi78,walt68}.

\begin{figure}
\includegraphics[width=8cm]{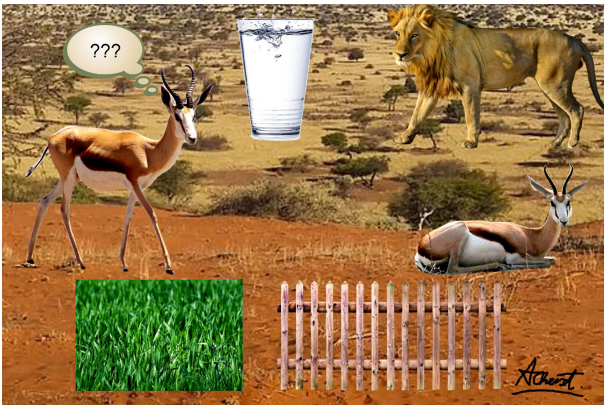}
\caption{Illustration of some vital daily decisions of a springbok, to choose the best survival, foraging, resting, etc. strategies of movement. Courtesy  www.pixabay.com for the source images and to Alexey A. Cherstvy for preparing the artwork.}
\label{fig_toc}
\end{figure}

In the current study, female \textit{Antidorcas marsupialis} (springboks) were
equipped with e-Obs GPS collars \cite{heringdon} for tracking. These animals
have a shoulder height of approximately 80 cm and weight between 30 to 40 kg.
They can reach speeds of up to 88 km/h in extended gallop \cite{bigalke1972}.
An image of a springbok together with its decision options is shown in
Fig.~\ref{fig_toc}.

Each collared
individual was selected from a different herd during the dry season and can
thus be treated to move independently from other tracked individuals. The
study area was located between the regions of Kunene, Omusati, and Oshana in
the north of Namibia, at $15.2235^\circ\mathrm{E}$, $19.2576^\circ\mathrm{S}$, approximately 80 km south-west of the Etosha pan, at Etosha Heights Private
Reserve and Etosha National Park. The vegetation zones in Etosha Park are known
to be very diverse, depending on the soil properties and water abundance.
Rainfall in the study area is highly variable, but mainly occurs from October
to April. During this wet season the mean daily temperature is around $26^
\circ\mathrm{C}$, with daily variations of some $15^\circ\mathrm{C}$. The dry
season is somewhat colder, with mean temperatures around $18^\circ\mathrm{C}$
and daily variations of around $20^\circ\mathrm{C}$. Our data set contains the
positions of eight springboks taken at time intervals of $\Delta t=15$ min for up to
31 months. The focal landscape is confined by a fence, which happens to be
damaged at places and thus allows some animals to cross or jump over it.
Detailed information on how animals interact with fences and on their energy
expenditure is available for female \textit{Antidorcas marsupialis} (springbok),
\textit{Tragelaphus oryx} (eland), and \textit{Tragelaphus strepsiceros} (kudu)
\cite{heri22fence}.

\begin{figure*}
(a) \includegraphics[width=8.2cm]{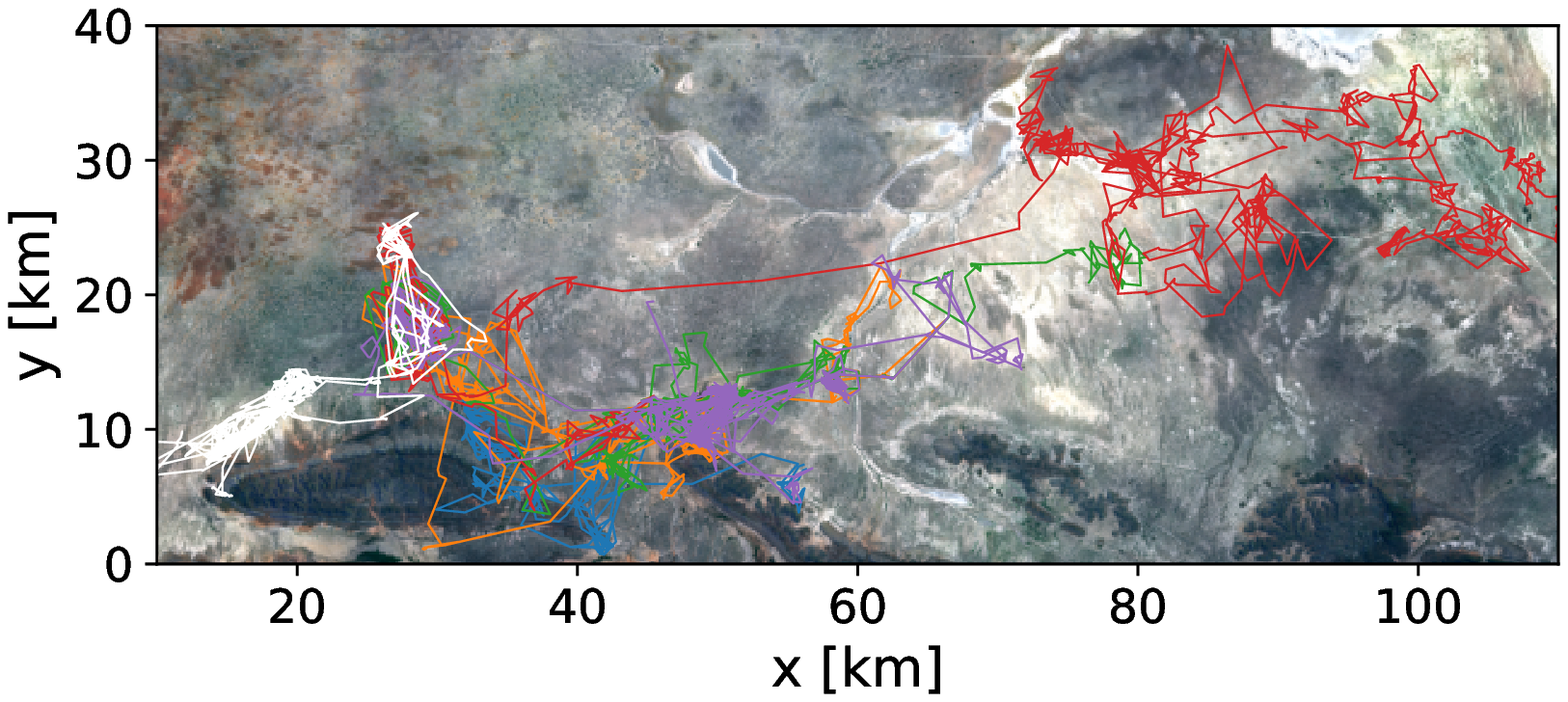}
(b) \includegraphics[width=8.2cm]{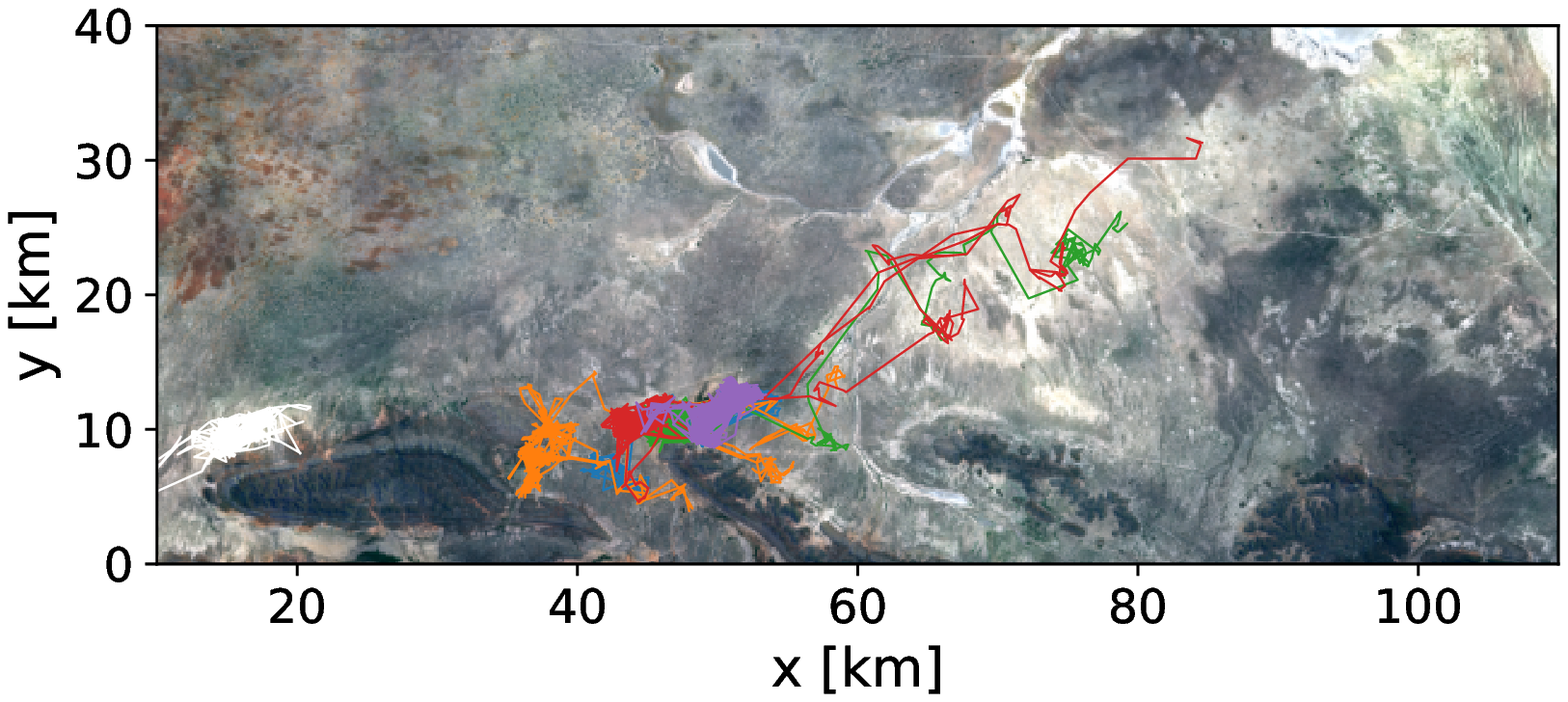}
\caption{Movement data of all springboks in the data set during (a) the wet
season of 2019/2020 and (b) the dry season of 2020. Different colors denote
distinct animals.}
\label{fig_data}
\end{figure*}

The average precision of the GPS positioning of the GPS-tagged springboks was
quite high because the weather conditions and the vegetation structure were
ideal for satellite reception. Each single position in the data set corresponds
to an average of a sequence of five GPS records taken at one second intervals.
When not moving (one GPS-sensor was tracked while laying in the field) the
apparatus yielded an accuracy of about $\delta x_{\mathrm{err}}\approx2\mathrm{
m}$ in two-dimensional position acquisition. Some pre-processing of the data was
conducted. Thus, missing data points (if only up to one hour of data was missing)
were replaced with the previous values of the animal's position. Position data
were stored along with the underlying vegetation pattern from the data set and
with the recorded ambient temperatures.

\begin{figure*}
(a) \includegraphics[width=5.6cm]{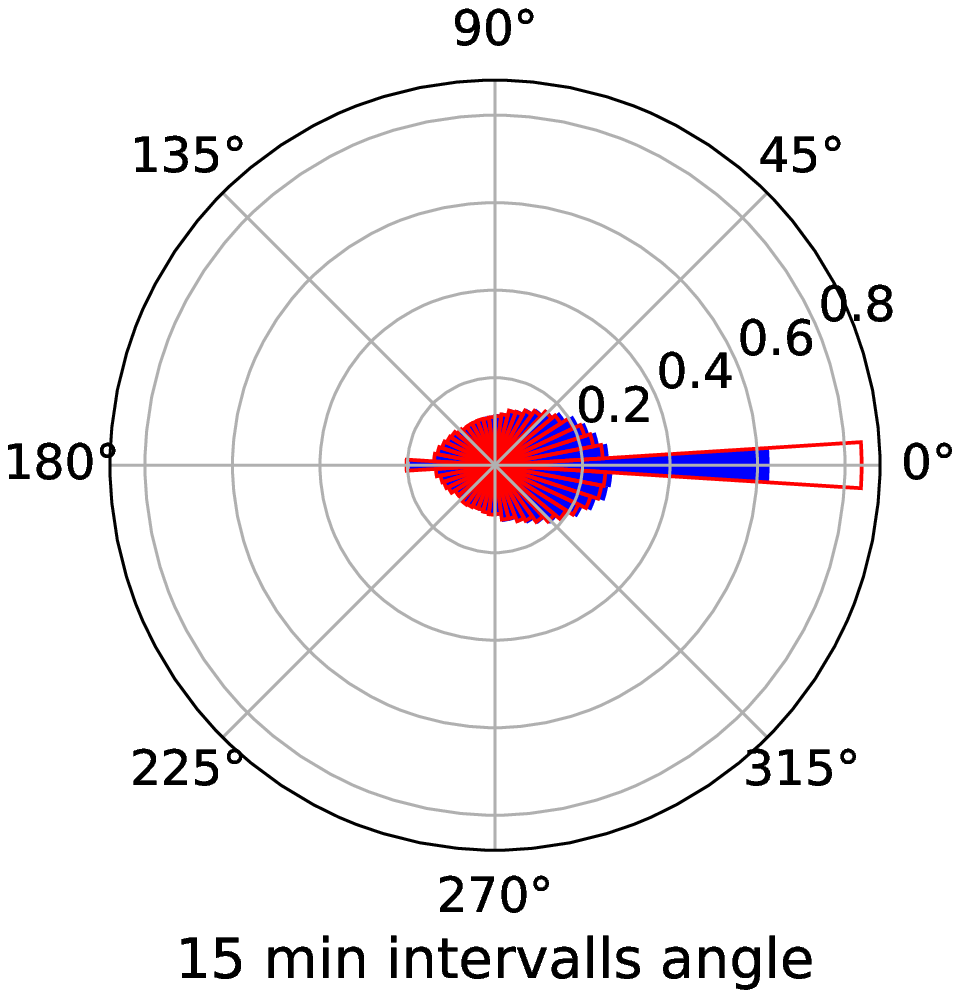}
(b) \includegraphics[width=5.6cm]{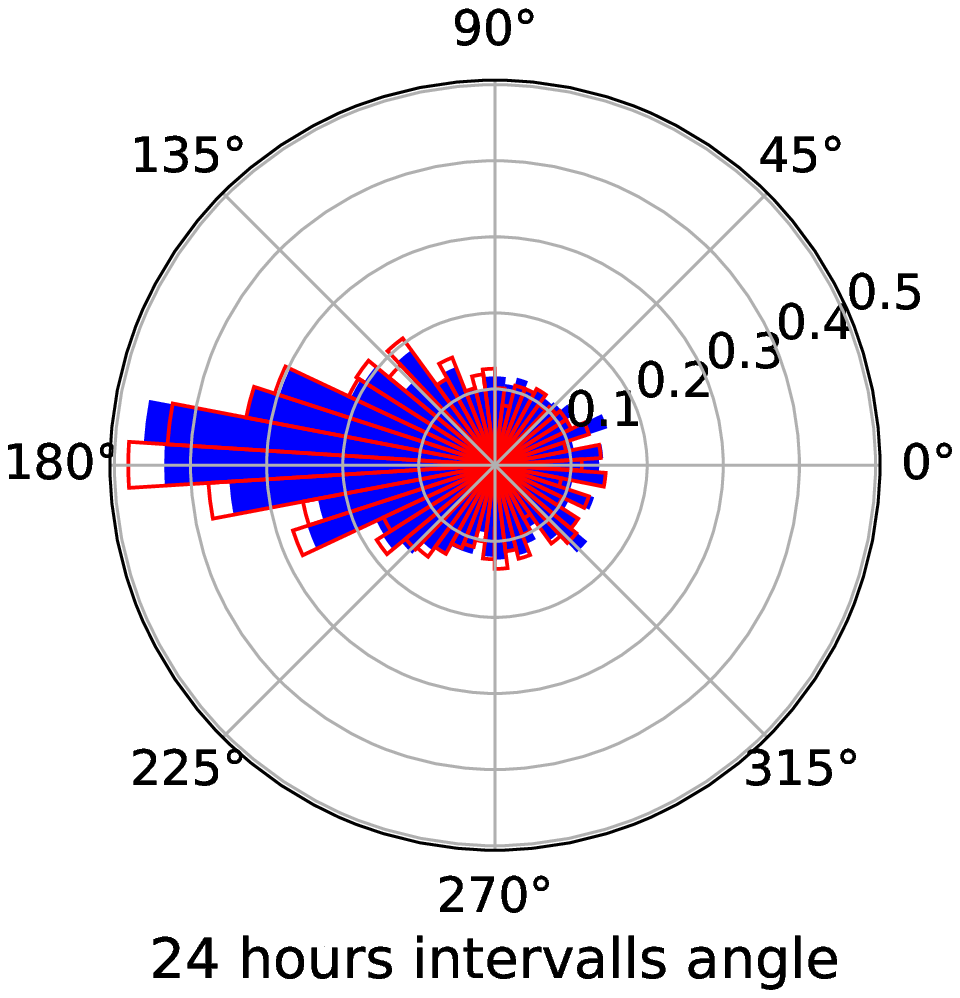}
\caption{Histograms of the turning angles between two successive steps of
springbok movement (data from all individuals), for data processed with (a)
15 min time steps and (b) one-day time steps. The full dataset (blue fill) is
shown together with the dry season data only (red outline).}
\label{fig_data_angles}
\end{figure*}

Springbok movement data during the wet and dry seasons are displayed in
Fig.~\ref{fig_data}. We observe that during the dry season the motion of
the animals is more localized around the water points, as the environmental
conditions necessitated regular returns to the water points for re-hydration.
The statistic of consecutive turning angles of all tracked springboks is
illustrated in wind rose diagrams in Fig.~\ref{fig_data_angles}. Visually
(see below for details) short-time persistent motion in the same direction
on 15 min intervals is distinct. Quite pronounced antipersistence is seen
on the daily time scale, signifying the eventual return to some preferred
location every day.

\section{Stochastic modeling}
\label{sec_tsd}

There exist a substantial number of both theoretical and data-based studies
dealing with animal motion modeling, see, e.g., \cite{BOVET1988419,Gurarie09}.
Typical observables are the moments of the motion or the corresponding position
or displacement autocorrelations. Correlations (persistence) in the motion is an
expected feature for directedness, and it may be associated as some measure for
the intelligence of the forager \cite{bhat2022smart}. Single trajectory power
spectral analyses may be additional quantities for analysis \cite{prx,vilkjpa}.
To describe the observed two-dimensional springbok motion we start with a
simplistic two-dimensional Brownian motion, that we write in the discrete form
\begin{equation}
\mathbf{r}(t_i)=\mathbf{r}(t_{i-1})+\xi(t_i)\mathbf{e}_\varphi(t_i),
\label{dtlan}
\end{equation}
where $\mathbf{e}_\varphi(t_i)=(\sin(\varphi(t_i)),\cos(\varphi(t_i)))$ and $t_i=i\Delta t$ is the time expressed in terms of elementary time steps
$\Delta t$, see below. The Gaussian driving longitudinal noise $\xi$ is of
zero mean and has the autocovariance function (ACVF) $\langle\xi(t_i) \xi(t_j)
\rangle=2K\delta_{ij}\Delta t$, where $\delta_{ij}$ is the Kronecker
delta.\footnote{Sometimes the discrete noise is also expressed in terms of
the "discrete-time impulse" $\delta[i-j]=\delta_{ij}$.} Moreover, $K$ is
the diffusion coefficient. The noise impulse $\xi(t_i)$ is allocated to the
two Cartesian co-ordinates via the random angle $\varphi(t_i)$, assumed to be
uniformly distributed on $[0,\pi)$. The MSD of this process then becomes
\begin{equation}
\langle\mathbf{r}^2(t_i)\rangle=4Kt_i,
\end{equation}
with $|\mathbf{r}(t_i)|
=\sqrt{x^2(t_i)+y^2(t_i)}$ and the initial condition $x(0)=y(0)=0$. The
angular brackets $\langle\cdot\rangle$ denote averaging over realizations of the
noise $\xi$. We note that on a log-log scale the MSD thus has unit slope.
From an individual time series $\mathbf{r}(t_i)$ of $N$ steps we calculate the
time-averaged MSD (TAMSD)
\begin{equation}
\overline{\delta^2(\Delta_l)}=\frac{1}{N-l}\sum_{i=1}^{N-l}\left(\mathbf{r}(t_{
i+l})-\mathbf{r}(t_i)\right)^2,
\end{equation}
in terms of the "lag time" $\Delta_l=l\Delta t$ \cite{pt}. As Brownian motion
is self-averaging, when $l\ll N$ we have
\begin{equation}
\label{bmtamsd}
\overline{\delta^2(\Delta_l)}=4K\Delta_l,
\end{equation}
that is, the process is ergodic in the Boltzmann-Birkhoff-Khinchin sense
\cite{pt,pnas}. In the following we drop the indices for discrete times, for
convenience.

\begin{figure*}
\begin{minipage}{0.35\textwidth}
(a)\includegraphics[width=0.95\textwidth]{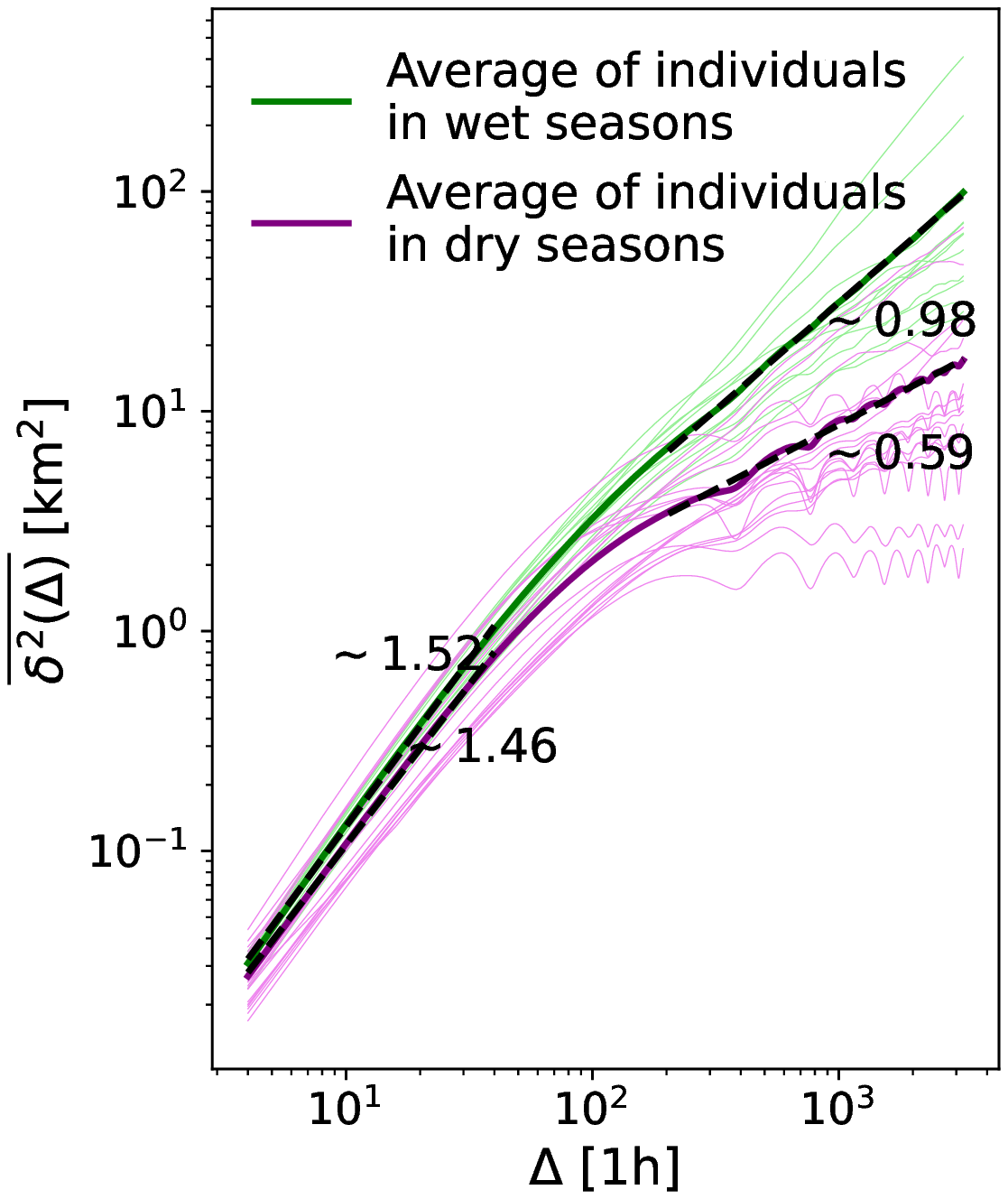}
\end{minipage}
\begin{minipage}{0.64\textwidth}
(b)\includegraphics[width=0.45\textwidth]{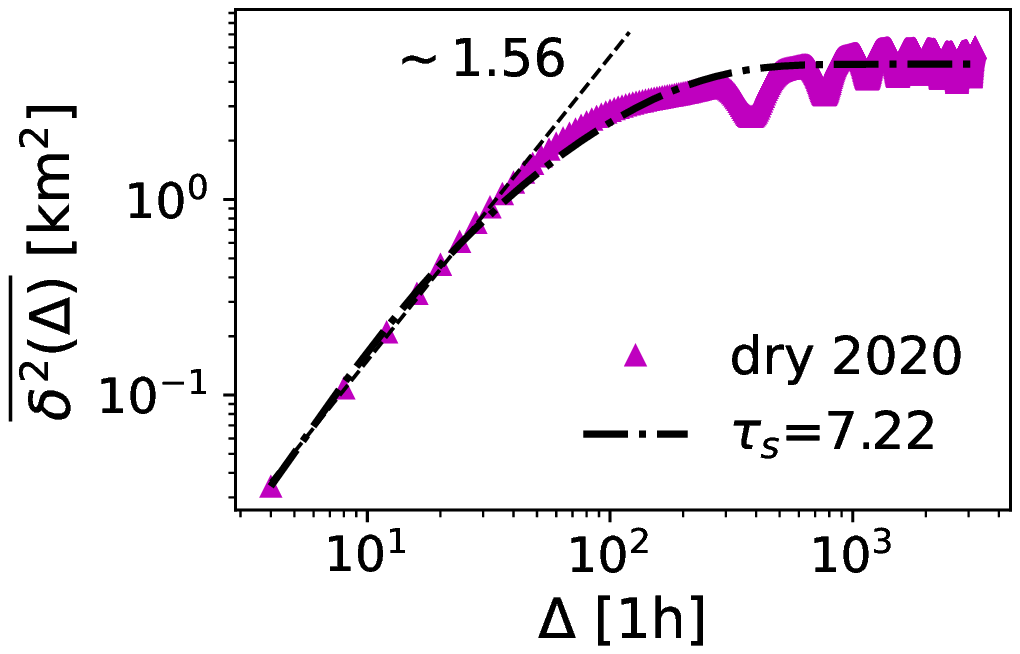}
(c)\includegraphics[width=0.45\textwidth]{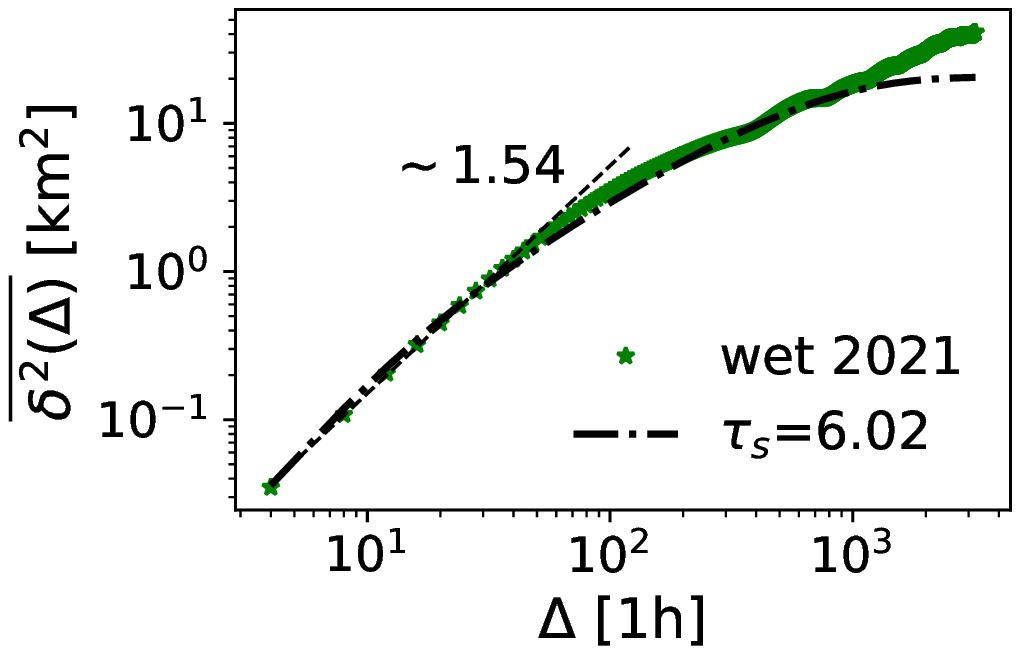}\\
(d)\includegraphics[width=0.45\textwidth]{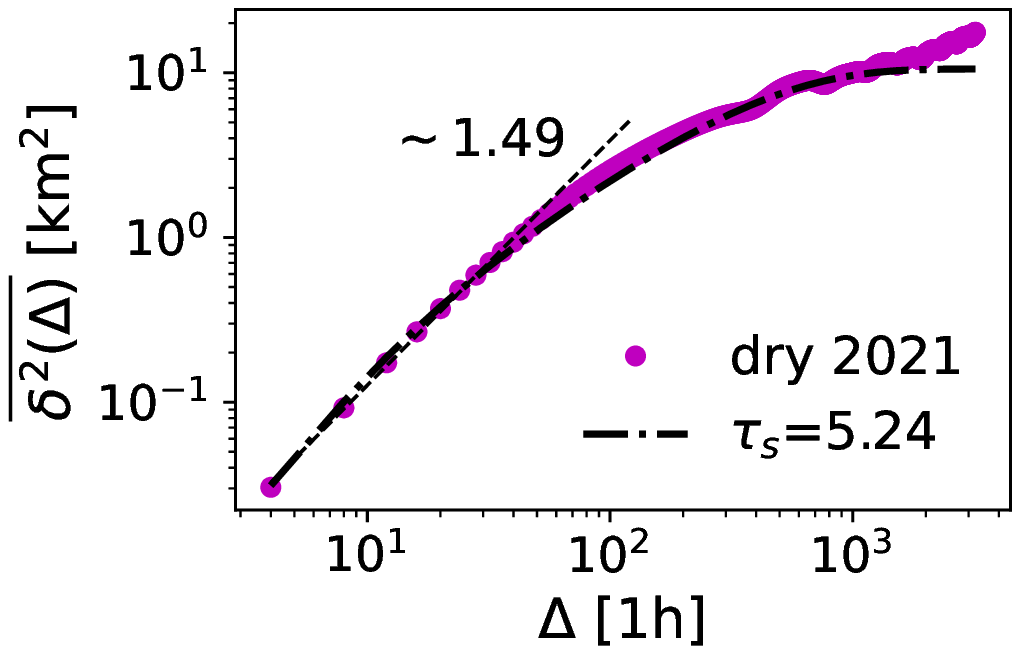}
(e)\includegraphics[width=0.45\textwidth]{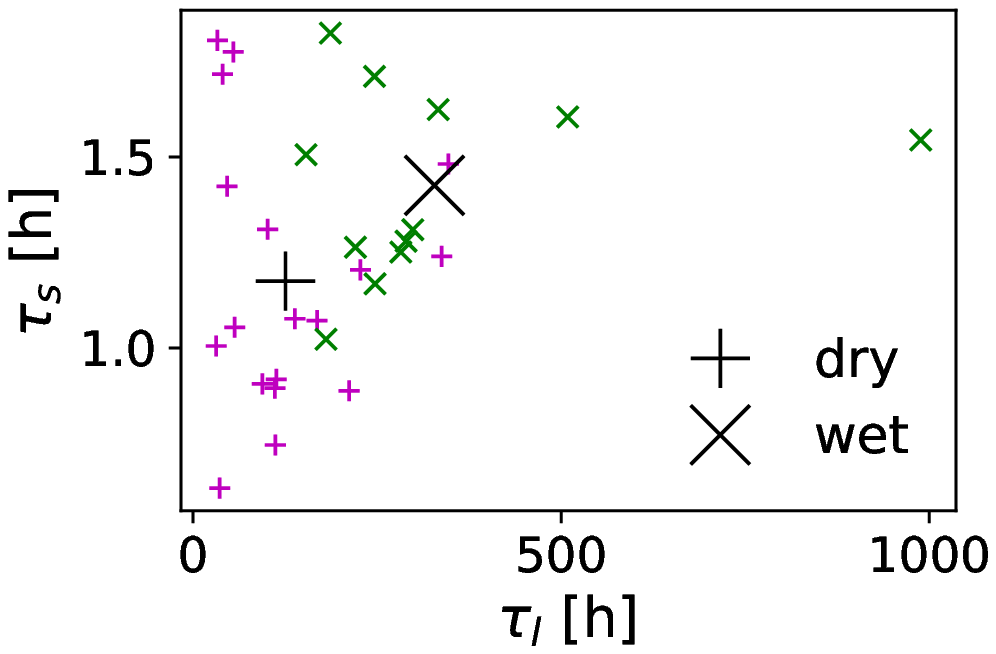}
\end{minipage}
\caption{(a) TAMSDs of springbok movement data with time steps of 15 minutes
for all individuals in all seasons. The thick lines are averages for all
TAMSDs from the the wet season and the dry season, respectively. The traces of
each animal were divided into wet and dry seasons, so that the total number of
(partial) trajectories is about three times larger than the number (8) of
individual animals in the data base. The slopes of the average TAMSDs for small
and large $\Delta$ are indicated by the fitted scaling exponents. The
oscillations in the TAMSD observed at intermediate-to-long lag times with a
period of 24 h are due to repetitive returns to certain preferential points,
such as favorite water and/or resting points.
(b-d) TAMSDs for dry (b,d) and wet (c) seasons for the same
individual; the dashed line shows the small $\Delta$ scaling; the theoretical
expression (\ref{eq_modeltsd}) with free parameters $\tau_s$, $\tau_l$, and $K$
is fitted and shown as the dashed-dotted line: Note that the oscillations in
the dry-season TAMSD curve (b) at intermediate to long times are artefacts of
the daily cycle.  (e) Fitted model parameters: the short and long time scales
$\tau_s$ and $\tau _l$ are compared for different individuals during dry and
wet seasons, see the legend and Eq.~(\ref{eq_modeltsd}) for details. Both
characteristic times are longer during the wet season in comparison to those
during the dry season, the difference is more significant for $\tau_l$.}
\label{fig_tsd}
\end{figure*}

Individual experimental TAMSDs of a number of individual springboks are shown
in Fig.~\ref{fig_tsd}a for wet and dry seasons, respectively, along with the
averages for each springbok ensemble. We see that the initial slope is greater
than unity, reflecting superdiffusive motion. As we will see, this superdiffusion
is due to persistence in a given direction of motion. After lag times of some 
70 to 100 h  the slope of the TAMSD changes, corresponding to traveled distances
of some 1 to 2 km. While for a number of trajectories the TAMSD continues to grow,
with a larger slope $\alpha\approx 0.98$ during wet seasons as compared to
$\alpha\approx 0.59$ during dry seasons, the TAMSD for some other trajectories
flattens off to a plateau value at around 100 h. Such a plateau value
within the measured lag time window appears more often in dry than in wet
seasons. We note that even for those animals, whose TAMSD grows until the maximum
lag time displayed in Fig.~\ref{fig_tsd}a, the TAMSD eventually also reaches a
plateau a fortiori, as the habitat is finite. Moreover, we point out that even
for the data
points at shorter times (smaller distances) the contribution $\delta x_{\mathrm{
err}}^2$ to the TAMSD due to measurement error is negligible, and thus the
extracted scaling exponents of the TAMSD are meaningful. We conclude from these
observations that the simple model of Brownian motion with TAMSD (\ref{bmtamsd})
is insufficient to account for the data. Instead, we are seeking a model that
captures the initial superdiffusion $\overline{\delta^2(\Delta)}\simeq\Delta^
{\alpha}$ with $\alpha>1$, a crossover to a second scaling regime $\overline{
\delta^2(\Delta)}\simeq\Delta^{\alpha'}$, and a terminal plateau behavior,
$\overline{\delta^2(\Delta)}\simeq\mathrm{const}$.\footnote{Here and in the
following, the symbol $\simeq$ denotes asymptotic scaling neglecting constant
coefficients.}

Let us first address the confinement effect. As an approximation we assume that
the animal motion is subject to an harmonic potential. Such a "soft" confinement,
in contrast to "hard" confinement in a finite box or in higher order potentials
such as $x^4$ forms, allows for variations in the maximal traveled distance.
These variations may occur, for instance, when the animal ventures out further
when a fence is broken or during periods of lusher vegetation. Discrete Brownian
motion in an harmonic confinement is then described by our discrete-time Langevin
equation (\ref{dtlan}) with {damping} coefficient $\exp(-\Delta t/\tau_l)$
\cite{ghosh,miao,qin},
\begin{equation}
\mathbf{r}(t_i)=e^{-\Delta t/\tau_l}\mathbf{r}(t_{i-1})+\xi(t_i)\mathbf{e}
_\varphi(t_i),
\label{eq_ou}
\end{equation}
This formulation is equivalent to the autoregressive model AR(1) of order one
\cite{Box15}, representing the discrete version of the seminal Ornstein-Uhlenbeck
(OU) process \cite{orns17,uhle30,risken,cher18ou}. The time scale $\tau_l$ in
the exponential prefactor in Eq.~(\ref{eq_ou}) is the characteristic correlation
time induced by the harmonic potential. At short times, the TAMSD of the process
(\ref{eq_ou}) is linear in time, while at long times the TAMSD converges to
$4K\tau_l$ \cite{meyer2022ho}.

In the short-time limit, the OU process with its linear scaling in time is thus
not an appropriate model for the observed springbok movement. Instead, animal
movement is characterized by a certain degree of persistence, i.e., the trend
to keep moving in a given direction \cite{romanczuk2012,gv}, as indicated by the
angle histograms in Fig.~\ref{fig_data_angles}. There are different modeling
approaches in literature for such persistence. As mentioned above, random search
by animals is often described by L{\'e}vy flights or walks. Due to the long
tailed jump length distributions, they perform superdiffusion. L{\'e}vy flights
in harmonic potentials lead to mono-modal (with a single maximum at the origin)
stable PDFs \cite{sune}, while in steeper than harmonic potentials their PDFs
exhibit bi-modal structures (with maxima away from the origin) \cite{chech,
chech1,karol}. L{\'e}vy walks with long-tailed jump distributions but a finite
propagation speed exhibit bi-modal PDFs already for harmonic confinement
\cite{pengbo}, including scenarios, in which the harmonic potential is only
switched on stochastically ("soft resetting") \cite{pengbo1}. Alternatively,
superdiffusive animal motion can be described by fractional Brownian motion
(FBM) \cite{vilk}, defined in terms of a Langevin equation driven by power-law
correlated Gaussian noise \cite{mandelbrot}. Positively correlated FBM in
steeper than harmonic potentials also exhibits bi-modal PDFs \cite{tobias,
tobias1}.

Here we will use a minimal model to introduce confinement and persistence
without long-range jump length distributions of power-law displacement
correlations. Namely, we will show that the springbok movement data can be
nicely described by replacing the white noise $\xi$ in Eq.~(\ref{eq_ou})
by exponentially correlated noise $\mathbf{z}$ \cite{maggi,fodor},
\begin{equation}
\mathbf{r}(t_i)=e^{-\Delta t/\tau_l}\mathbf{r}(t_{i-1})+\mathbf{z}(t_i),
\end{equation} 
where we choose
\begin{equation}
\mathbf{z}(t_i)=e^{-\Delta t/\tau_s}\mathbf{z}(t_{i-1})+\sqrt{\frac{
\Delta t}{\tau_s}}\xi(t_i)\mathbf{e}_\varphi(t_i).
\end{equation}
Here $\xi$ is white Gaussian noise and $\tau_s$, chosen as $\tau_s\ll\tau_l$,
is the correlation time of the driving noise $\mathbf{z}$. Comparing with
Eq.~(\ref{eq_ou}) it is clear why the noise $\mathbf{z}$ is often called OU noise.
These equations can be simplified by eliminating $\mathbf{z}$, yielding the
autoregressive process AR(2) of order two \cite{Box15},
\begin{eqnarray}
\nonumber
\mathbf{r}(t_i)&=&(e^{-\Delta t/\tau_l}+e^{-\Delta t/\tau_s})\mathbf{r}(t_{i-1})\\
&&-e^{-\Delta t/\tau_l-\Delta t/\tau_s}\mathbf{r}(t_{i-2})+\sqrt{\frac{\Delta t}{
\tau_s}}\xi(t_i)\mathbf{e}_\varphi(t_i).
\label{eq_model}
\end{eqnarray}

The TAMSD defined by the model (\ref{eq_model}) reads \cite{meyer2022ho}
\begin{equation}
\label{eq_modeltsd}
\left<\overline{\delta^2(\Delta)}\right>=\frac{4K\tau_l^2}{\tau_l^2-\tau_s^2}
\left[\tau_l\left(1-e^{-\Delta/\tau_l}\right)-\tau_s\left(1-e^{-\Delta/\tau_s}
\right)\right].
\end{equation}
In the short time limit $\Delta\ll\tau_s$, this expression encodes the
ballistic scaling
\begin{equation}
\left<\overline{\delta^2(\Delta)}\right>\sim\frac{2K\tau_l}{\tau_s(\tau_l
+\tau_s)}\Delta^2,
\end{equation}
whereas at intermediate times $\tau_s\ll\Delta\ll\tau_l$ we find a linear
$\Delta$-dependence with a correction term,
\begin{equation}
\left<\overline{\delta^2(\Delta)}\right>\sim\frac{4K\tau_l^2}{\tau_l^2-\tau_s^2}
\left(1-\frac{\Delta}{\tau_l}\right)\Delta.
\end{equation}
At long times, $\Delta\gg\tau_l$, the TAMSD reaches the plateau value
\begin{equation}
\left<\overline{\delta^2(\Delta)}\right>\sim\frac{4K\tau_l^2}{\tau_l+\tau_s}.
\end{equation}
The characteristic times $\tau_s$ and $\tau_l$ describe the dominant
dependencies of the TAMSD in the limts of short and long times, respectively
(as indicated by the indices). We use the TAMSD to assess the typical
ballistic and confined motion (converging to a plateau) of sprtingboks expected
at short and long times.

The solution (\ref{eq_modeltsd})
can be fitted to the measured TAMSD of the springbok data. As we can
see in Fig.~\ref{fig_tsd}b the agreement is quite good. At shorter
times we see that the model with the three fit parameters $K\sim
1\ldots10\mathrm{km}^2/\mathrm{h}$, $\tau_s\sim 1\mathrm{h}$, and
$\tau_l\sim1\mathrm{day}$, matches the data well up to the time scale of a
week, and levels off to the plateau somewhat too early. However, given that
the initial power-law regime spans merely around one decade, this does not
appear too severe. In contrast, we believe that the relatively simple AR(2)
model allows for easy physical interpretation of the parameters and provides
a very satisfying description (for more details see the data sections
below). Importantly, in this AR(2) model we can easily include an error
analysis relevant for experimental data.

Before we continue we briefly describe our fitting procedure of TAMSD
curves as those shown in Fig.~\ref{fig_tsd}b. We use a nonlinear fit of the
TAMSD (\ref{eq_modeltsd}), equivalent to a nonlinear fit to the fluctuation
function \cite{meyerDFA}.  The model has three free parameters $K$, $\tau_l$,
and $\tau_s$. The fact that the most reliable points of the TAMSD are those
at shorter lag times while most points are available at longer times (due to
the logarithmic scaling) makes the fit challenging. The solution we choose is
to fit one parameter at a time. For the fit shown in Fig.~\ref{fig_tsd}b we
therefore practically divide the lag time window into separate intervals and
fit the time scales to the relevant range of lag times. This is possible as
long as $\tau_l\gg\tau_s$, which we obtain from the data self-consistently. We
start measuring the variance $\sigma^2$ (related to the diffusivity $K$) of
the time series.  As the curvature of the TAMSD is independent of $\tau_s$
for $\Delta\gg\tau_s$, we set $\tau_s$ to a small value and fit $\tau_l$
in the range between $1.25$ h and 3.75 h. Finally, $\tau_s$ is fitted using
the parameters $K$ and $\tau_l$, and the first two points of the dataset.

The results are displayed in Fig.~\ref{fig_tsd}. In panel (a) we show all
seasons and individuals along with the average for each season. Panels
(b-d) are examples from one individual during three different seasons.
In Fig. \ref{fig_tsd}c we show all measured parameters $\tau_s$ and $\tau_l$
of different animals during movement in both wet and dry seasons.  Generally,
$\tau_l$ is larger in the wet season, reflecting a confinement of animals
due to a lack of resources during the dry season.  The time scale $\tau_s$
also has a tendency to be smaller in dry seasons. There is, however, a large
overlap in the found distributions of $\tau_s$ and $\tau_l$ when comparing
seasons. In all situations the strong inequality
\begin{equation}
\tau_l\gg\tau_s
\end{equation}
holds, so that in the short-time limit the dynamics of animals can be
approximated by free diffusion with correlated driving $\zeta$.

\begin{figure*}
\centering
a)\includegraphics[width=0.31\textwidth]{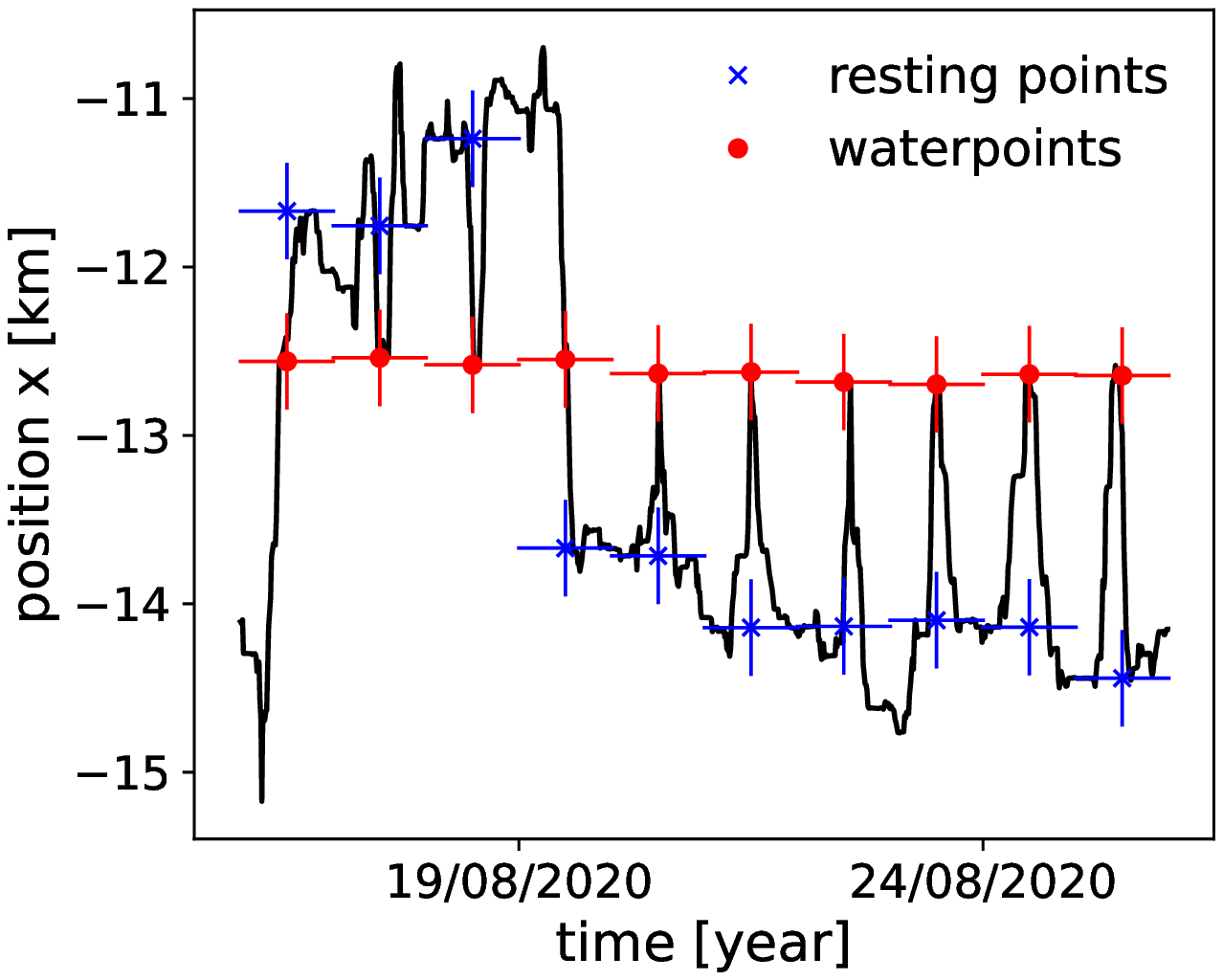}
b)\includegraphics[width=0.31\textwidth]{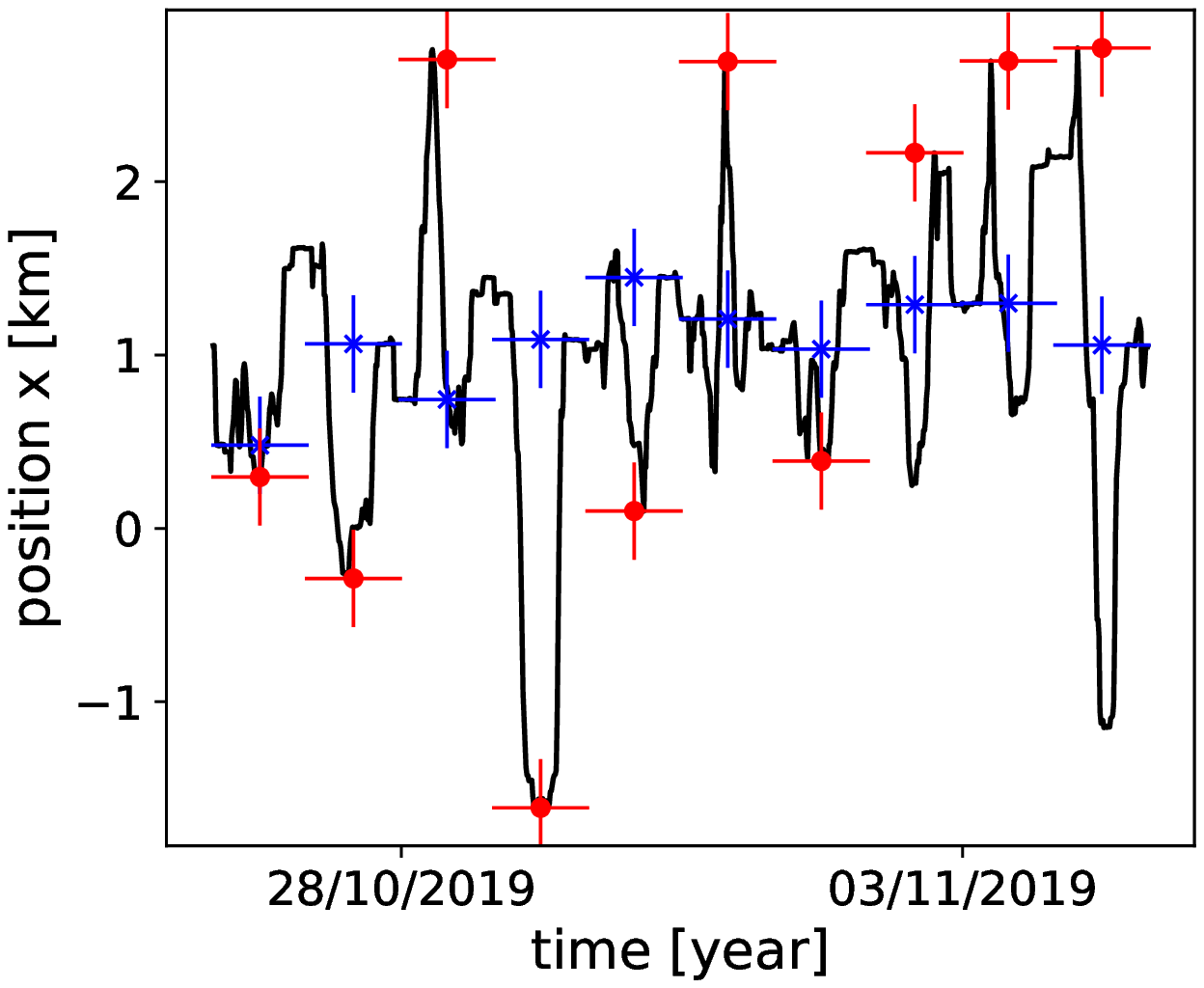}
c)\includegraphics[width=0.31\textwidth]{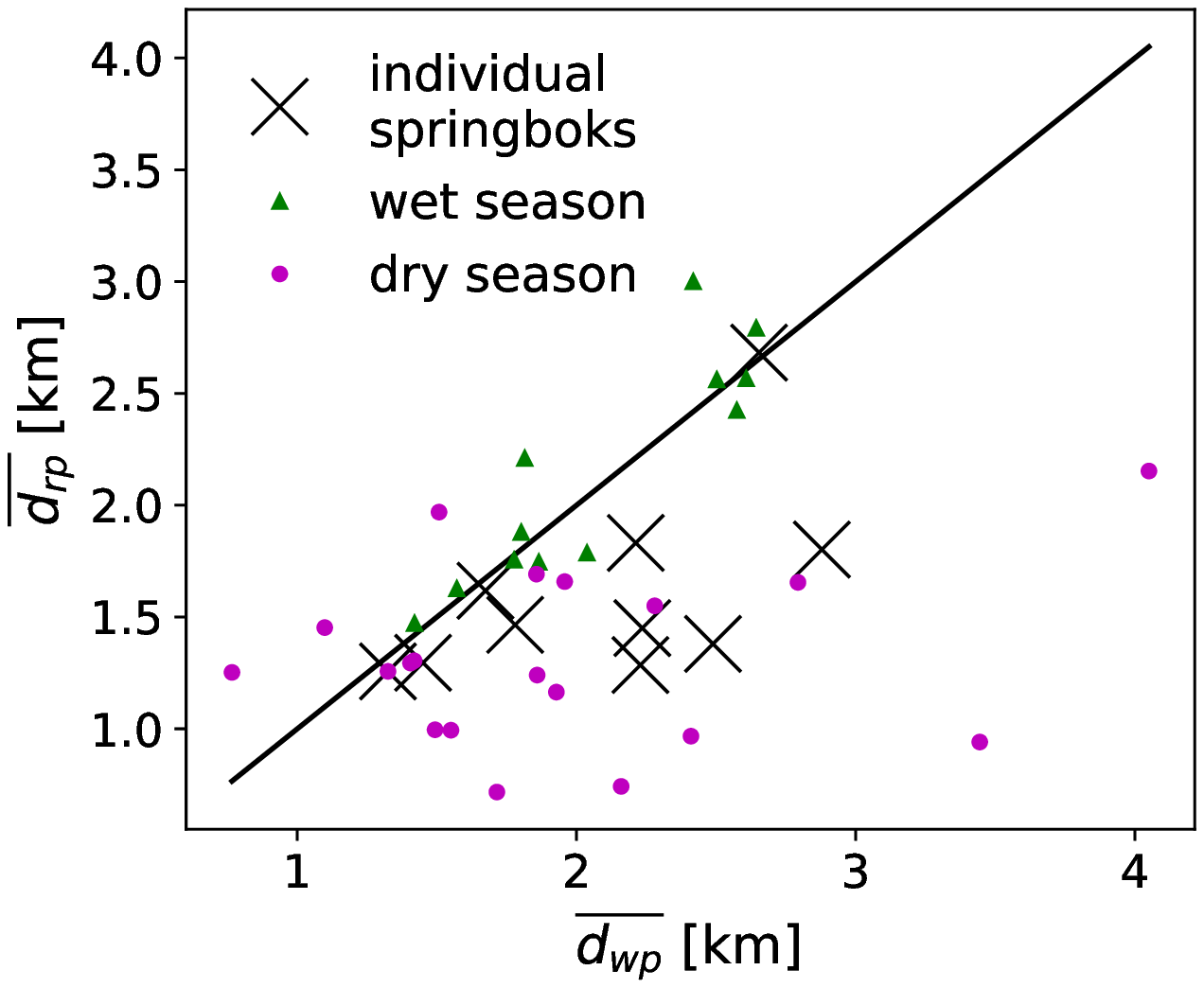}
\caption{Effects of water and resting points. (a) Segment of the trajectory
$x(t)$ (position time series projected onto the $x$ direction) during which the
springbok visits the same water point every day, while the maximum distance
and the resting points vary. (b) Segment of another trajectory, during which
the resting point remains almost constant while the animal visits different
water points. (c) Mean distance between the resting points on two subsequent
days versus the distances of visited water points on subsequent days. Large
crosses are calculated for the full data set of single individuals. The small
symbols are from the same data sets, but averages are taken only over the 6
months of either dry or wet season, see legend. The black line represents
the diagonal.}
\label{fig_wpsp}
\end{figure*}

\section{Feature characteristics}
\label{sec_param}

The above model takes into account finite-time correlations in the movement
and confinement. We here discuss the influence of two important additional
features, namely, geographical features such as water and resting points,
and temporal features such as day and night.

\subsection{Water and resting points}

Water points are a key element in springbok movement dynamics. Although
springboks are well adapted to arid environments and can survive long periods
without drinking \cite{skin96}, they regularly visit water points when
available.  This way, they avoid negative physiological consequences during
the adjustment to a low water supply \cite{Cain2006spwater}. Springboks drink
at water points throughout day time but also occasionally during night time
\cite{bigalke1972}.  To understand the springbok movements it is necessary
to investigate the effects of water points.

During the wet season, when there is ample nutritious food available, it is
known that large herds of hundreds of springboks concentrate on open, highly
productive grassland \cite{bigalke1972}. During the dry season, when food is
more limited, the springboks disperse and form smaller groups of a few dozen
of individuals \cite{bigalke1972}. Such behavior may, however, vary between
different areas \cite{davi78,Roche2008} and largely depends on the combined
environmental factors such as water point location and their water levels,
fencing, general geo-hydrology, weather, and other. The behavior with respect
to the water holes is not unique, as shown in Fig.~\ref{fig_wpsp}a,b. The
segment of the animal trajectory shown in panel (a) shows a preferential
return to the same water point while different resting points are chosen
and the animal has a fairly constant maximal range of motion on consecutive
days. In contrast, in the trajectory segment in panel (b) the same individual
covers much smaller daily distances (roughly 10\% of the covered spans of
panel (a)), and shows only small variations in the resting points, while it
visits continuously changing water points.

The springbok resting points at night correspond to the maximum of their
temporal occupation density, i.e., the local area where they spend most time in
a 24 hour span. In some periods, the resting points of animals are almost the
same every night, see Fig.~\ref{fig_wpsp}b, while in other cases the resting
points are more distant (panel (a)). In order to find the maximum of the
probability density of springbok positions on a specific day, one approach
is to recursively delete the point which is furthest away for the median.
This way we start with the full set $\{x(T),y(T)\}$ of positions of length
$24\times4$ for all recorded positions during the span of a single day ($T$
runs from midnight to 23:45 hours). We then enumerate the distance $([x(T)-
\mathrm{median}(x(T))]^2+[y(T)-\mathrm{median}(y(T))]^2)^{1/2}$. The elements
$\{x(T_\mathrm{max}),y(T_\mathrm{max})\}$ which has the maximal distance
from the median of $x$ and $y$ is then deleted, and the procedure is repeated
with the next point, until only one point is left in the set. This identifies
the maximum of the probability density.

To quantify the effects of water and resting points on the springbok movement, we
define, for each day, the point closest to a water point as $\{x_{\mathrm{wp}},
y_{\mathrm{wp}}\}$ and the resting point, at which the density of the visited
positions of the animal reaches a maximum, as $\{x_{\mathrm{rp}},y_{\mathrm{rp}}
\}$. This is a somewhat crude approximation, because springboks do not necessarily
visit a water point only once per day and do not have a single sleeping phase per
day. However, the two sets of data points $ \{x_{\mathrm{wp}},y_{\mathrm{wp}}\}$
and $\{x_{\mathrm{rp}},y_{\mathrm{p}}\}$ are, as we show, meaningful quantities
for analysis.

We compute the mean distance of subsequent daily water or resting points as
\begin{eqnarray}
\nonumber
d_{\mathrm{wp}}(T_i)&=&\Big([x_{\mathrm{wp}}(T_i)-x_{\mathrm{wp}}(T_i-1)]^2\\
&&+[y_{\mathrm{wp}}(T_i)-y_{\mathrm{wp}}(T_i-1)]^2\Big)^{1/2},
\end{eqnarray}
where $T_i$ is the sequence of days. Analogously, we evaluate $d_{\mathrm{rp}}
(T_i)$ for the resting points. The mean of these expressions for different
individuals in the tracked ensemble are compared in Fig.~\ref{fig_wpsp}c.
We find that, in general, the resting point distances vary less than those
of the water points do. The analysis separating wet and dry seasons shows
that the difference in variation is mostly due to the movement in the wet
seasons, while in the dry seasons both time series vary similarly. The fact
that longer journeys are necessary to find sufficient water supplies in the
dry season reflects the expected conditions in arid climates. On average,
therefore, we conclude that springboks rather keep to more localized resting
points and travel further for water points.

A video of the motion of a springbok on an actual map with a variable
vegetation index is presented in Fig.~\ref{video-pred}, together with
motion-step predictions from our theoretical model (see below). Note that
the modeled springbok is always somewhat lagging behind the actual measured
springbok positions because predictions are weighted averages over all possible
outcomes and, therefore, 2D-predictions tend to have somewhat smaller steps
as compared to the true step lengths.  The overall reproducibility of the
movement directions, the magnitude of the turning angles, and the "intensity"
of motion is, however, remarkable for the two springboks chosen from the dataset.

\begin{figure}
\centering
%\textattachfile[mimetype=video/mp4,description=double-click to load video in gif format,]{prediction.mp4}{\includegraphics[width=0.45\textwidth]{prediction.eps}}
\includegraphics[width=0.45\textwidth]{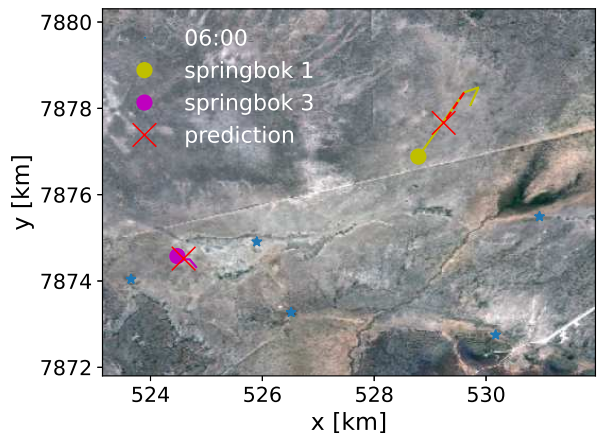}
\caption{Snapshot from a movie of the motion of two springboks: the water points
are marked as blue stars on the map, the actual animal positions are the orange
and violet filled circles. The model predictions (for the same time intervals
between steps) are denoted by the red crosses. The full movie file is provided
in the Supplementary Material \cite{sm}.}
\label{video-pred}
\end{figure}

\subsection{Day- and night-autocorrelation}

\begin{figure*}
a)\includegraphics[width=0.19\textwidth]{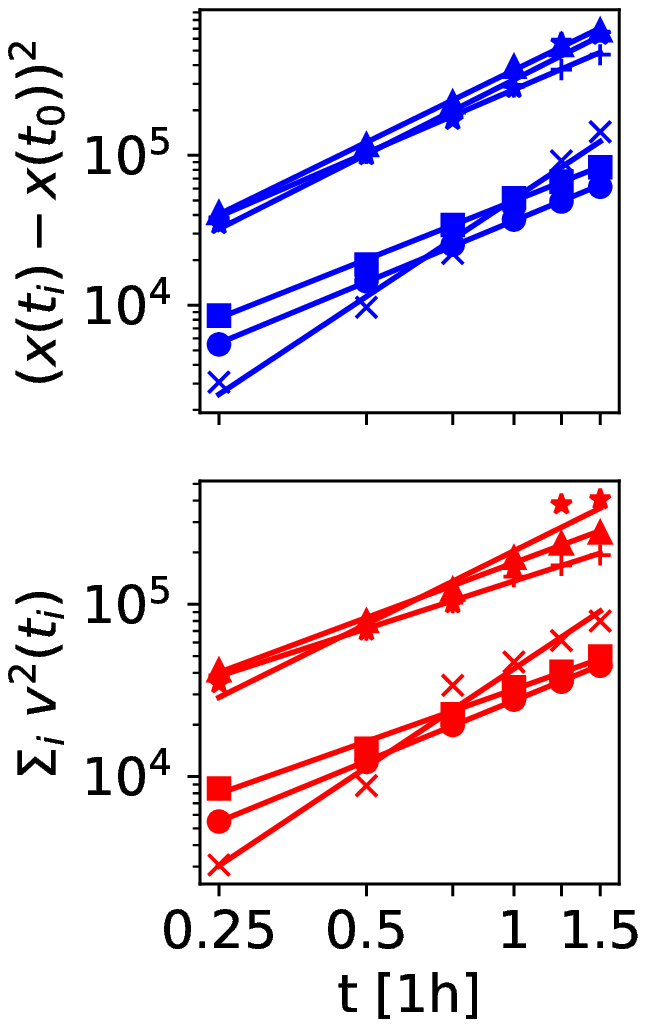}
b)\includegraphics[width=0.475\textwidth]{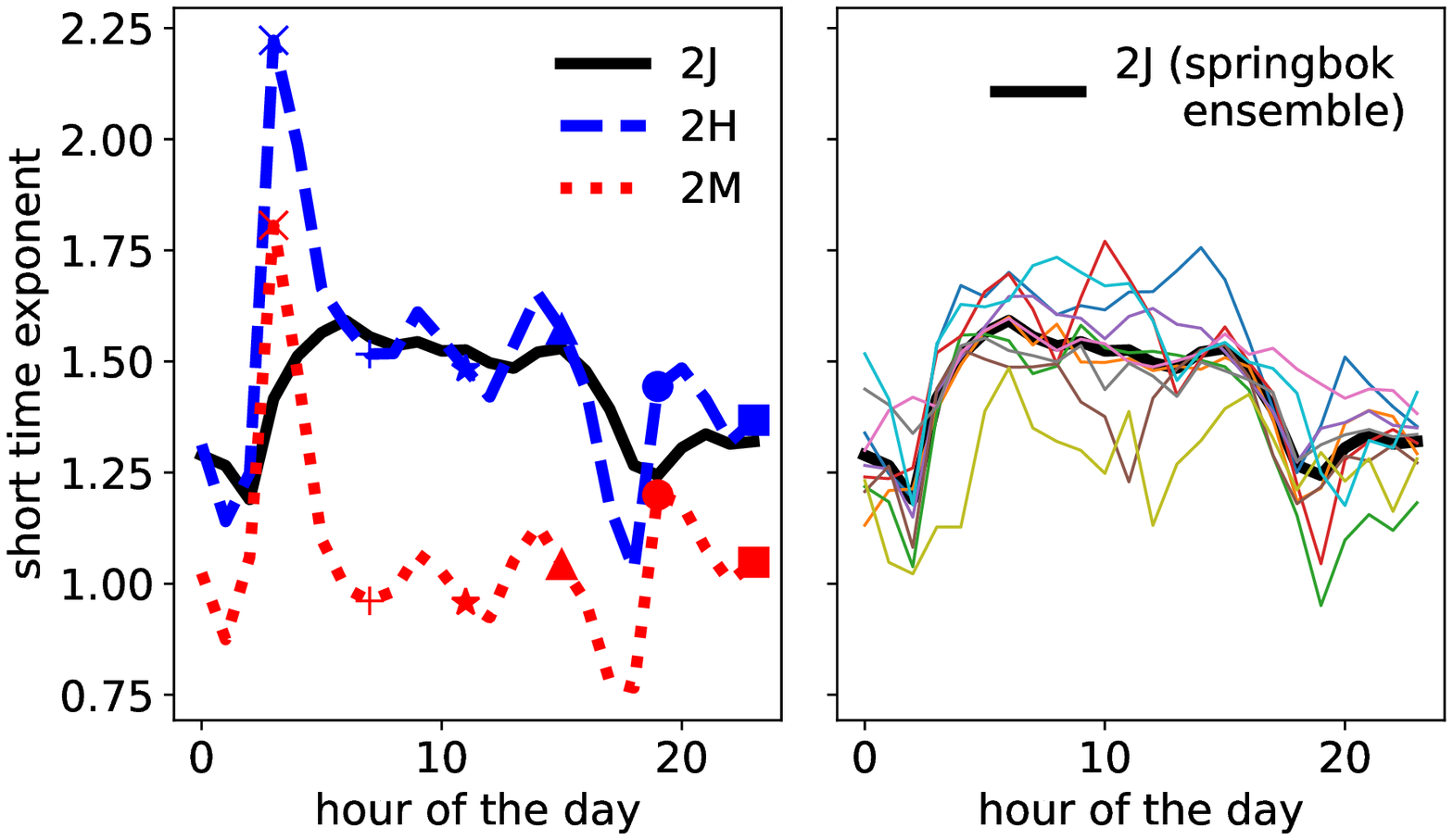}
c)\includegraphics[width=0.262\textwidth]{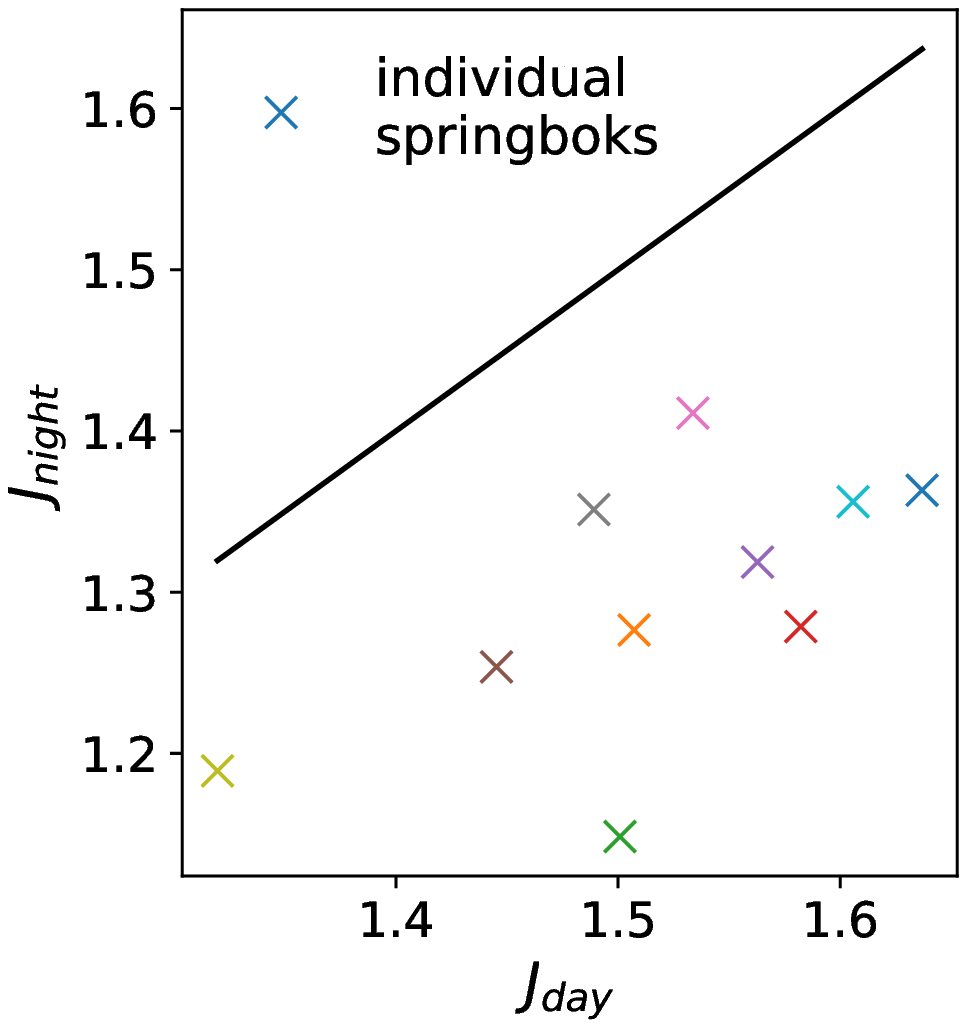}
\caption{Displacements and short time scaling exponents of the ACVF of
springbok motion. (a) Single springbok data, top: Scaling of the squared
displacements along the trajectory for different hours of the day; bottom:
cumulative sum of the squared velocities of springbok motion for different
hours of the day. The different symbols represent different starting hours. We
observe a spread in values of the scaling exponents depending on the time
of day.  (b) Springboks ensemble data, left: for each hour of the day,
the average scaling exponents $\{2J,2H,2M\}$ are calculated as an average
over days, using 15 min intervals. The symbols correspond to the ones used
in panel (a).  Exponents are disentangled following Eq.~(\ref{moses});
right: for the long-range correlation exponent $J$, thin lines represent
different individuals, the thick line is an average over all individuals.
(c) Correlations between the scaling exponents $J$ during the day and night.
Average scaling during the day (4 am to 6 pm) and night (6 pm to 4 am)
shows a variability among the individuals and an increased directedness
(i.e., more positive $J$  values) during the day for all animals in the
dataset. The black line represents the diagonal.}
\label{fig_dnac}
\end{figure*}

We now address the dynamics encoded in the movement data on the level
of a single day. The data segments shown in Fig.~\ref{fig_wpsp} suggest
some daily cycle regarding the distances covered by the springboks.
In this section we address the question whether this daily cycle in walked
distance is due to a daily cycle in activity or due to a daily cycle in
directedness of the motion or both.  To this end, we need to assess all
three quantities, the total distance, the activity and the directedness.
With changing activity, we mean non-stationarity of the driving, related to
the parameter $K$. Here, by directedness we mean short-time autocorrelations.
One difficulty in distinguishing these two characteristics is the fact that
our data has the resolution of only 15 min and thus shorter correlation
times cannot be resolved.  First, we need an assumption regarding the
shape of short-time autocorrelations prior to starting the analysis. From
Fig. \ref{fig_tsd}(a-d) it follows that a power-law shape is a reasonable
assumption for the short-time behavior of the TAMSD, which is closely related
to the autocorrelation function \cite{meyer18,meye21moses,meye22moses}. So,
for simplicity, we assume a power-law TAMSD scaling with a corresponding
exponent in the short-time limit of the displacement.  The total distance
walked by an individual can be measured by the squared displacement
$r^2_h(t)=(x_h(t)-x_h(0))^2+(y_h(t)-y_h(0))^2$, where the
index $h$ denotes the time of the day when the measurement is started, and
$r_h(0)$ is the starting position. One question is whether the
growth of the squared displacement depends on the starting time,
i.e. the hour $h$ of the day. Fig.~\ref{fig_dnac}a displays the average over
an ensemble of days in the data set (we assume that the days can be
considered independent). At each time of the day, we can assign a Hurst exponent
$H$ to the ensemble-averaged displacement $\langle r_h^2(t)\rangle$, compare
Eq.~(\ref{eq-msd-h-expon}). As described in App.~\ref{mosesapp}, we also define
scaling exponents for the velocity autocorrelation function (Joseph exponent
$J$), and of the non-stationarity (Moses exponent $M$). The latter can be
defined via the scaling of the cumulative squared increments of animals in $x$
and $y$ directions, i.e.
\begin{equation}
\label{eq-msd-m-expon}
\left<\sum_{i=1}^n\mathbf{v}^2(t_i)\right>\propto n^{2M+1},
\label{eq_M}
\end{equation}
with increments $\mathbf{v}(t_i)=[\mathbf{r}(t_i)-\mathbf{r}(t_{i-1})]/\Delta t$.
It can be shown that the exponents $H$, $M$, and $J$ are related to each other
via the following relation \cite{mand68moses,che17,meyer18,aghion2021moses,vilk} 
\begin{equation}
\label{moses}
J=H-M+1/2,
\end{equation}
as in our case due to the finite variance of the increments the Noah exponent
$L=1/2$ (see also App.~\ref{mosesapp}).

Given relation (\ref{moses}), it is sufficient to calculate two of the three
exponents for a certain hour of the day for the springbok motion to know
all three of them. As the inference of autocorrelations usually involves
time-averaging, we restrict ourselves to calculating the exponents $H$
and $M$.  Here $M$ is defined by relation (\ref{eq_M}) and $H$ is defined
by the ensemble-averaged MSD  \cite{meyer18,meye21moses,meye22moses},
\begin{equation}
\label{eq-msd-h-expon}
\left<(\mathbf{r}(t_i)-\mathbf{r}(t_0))^2\right>\propto(t_i-t_0)^{2H}.
\end{equation}
In Eqs.~(\ref{eq-msd-m-expon}) and (\ref{eq-msd-h-expon}) the brackets denote
averaging over different days of springbok data and for each day $t_0$
corresponds to the same hour of the day. Using segments of length 90 min,
we extract the short-time exponents.

In Fig.~\ref{fig_dnac}b,c we show the results of the computations for different
times of the day and for different springboks. Figure \ref{fig_dnac}b
demonstrates the evolution of the short-time scaling exponents $\{J,H,M\}$
for all individuals measured at each hour of the day. The exponents $H$ and
$M$ peak in the morning, reflecting the activity increase of
the springboks increases fast as the sun rises. Note, that the exact time of the
sunrise varies for each calendar day, but due to the location of springbok
trajectories close to the equator this effect is relatively minor.

In Fig. \ref{fig_dnac}c we plot the average exponent $J$---describing
the directedness of the motion---during the day (4 am until 6 pm) versus
the exponent computed during the night (6 pm until 4 am).  We find that
autocorrelations---quantified by the value of exponent $J$---during the day are
\textit{more} pronounced as compared to those during the night. During the
entire day, the motion is more persistent than a completely random process
($2J>1$), but less directive/persistent than a ballistic process ($2J<2$).

We stress here that the underlying assumption of the existence of the
power-law TAMSD scaling is a strong approximation and thus the inferred
exponents are to be treated with care. According to Fig. \ref{fig_tsd}b,
the TAMSD$(\Delta)$ has a concave shape and thus autocorrelations in
reality might somewhat higher at very short time-scales (that is
below the time resolution of the current analysis (assuming power-laws)).
The inferred influence of the autocorrelations can, therefore, be regarded
as a lower bound to the impact of autocorrelations.

We find that changes in both activity and directedness contribute to the
daily cycle. Both are not accounted for by model (\ref{eq_model}).

\section{Predictability}
\label{sec_pred}

The analysis in the previous Section demonstrates that our relatively simple
Ornstein-Uhlenbeck model (\ref{eq_model}) with correlated noise cannot fully
describe all features of the springbok movement dynamics. While this simple
model captures several essential features of the dynamic, it would
be naive to expect that such rather generic physics-inspired models could grasp
the complexity of effects such as day-night or seasonal variations. One
possible generalization could include different parameters for day and night.
The heterogeneity of the springbok movement stems from the environment or
terrain features, but also from different modes of individual animal movement
(such as, generally, hunting, resting, hiding, vigilance, etc.). Some studies
specifically focus on classifying these modes of motion
\cite{clintock2018hmm,patin2020segclust2d,bhat2022smart}.
Here we discuss some possible extensions of our model and estimate their
predictive power.

\subsection{Comparison with basic AR(1) and AR(2) models}

\begin{figure*}
a)\includegraphics[width=0.31\textwidth]{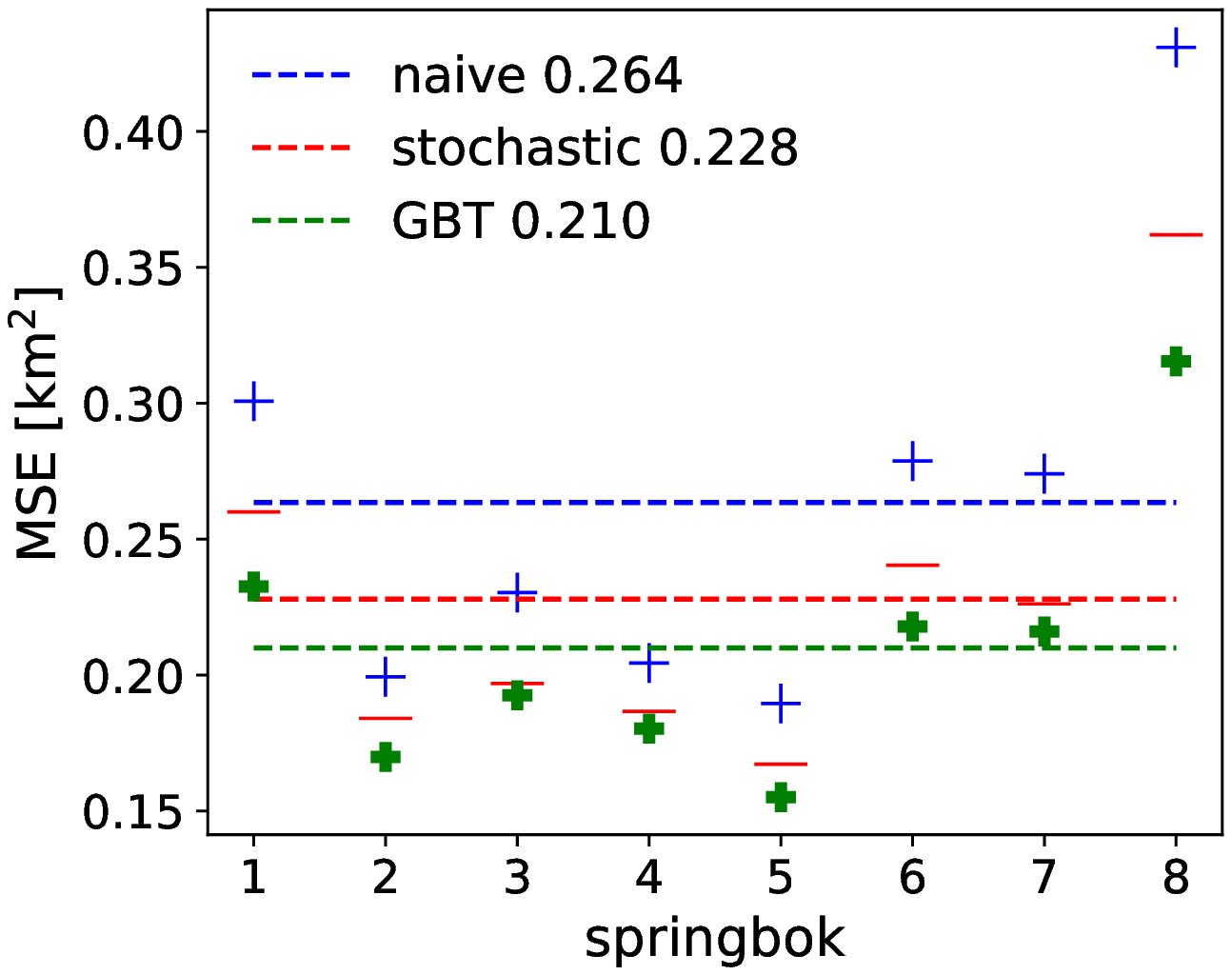}
b)\includegraphics[width=0.31\textwidth]{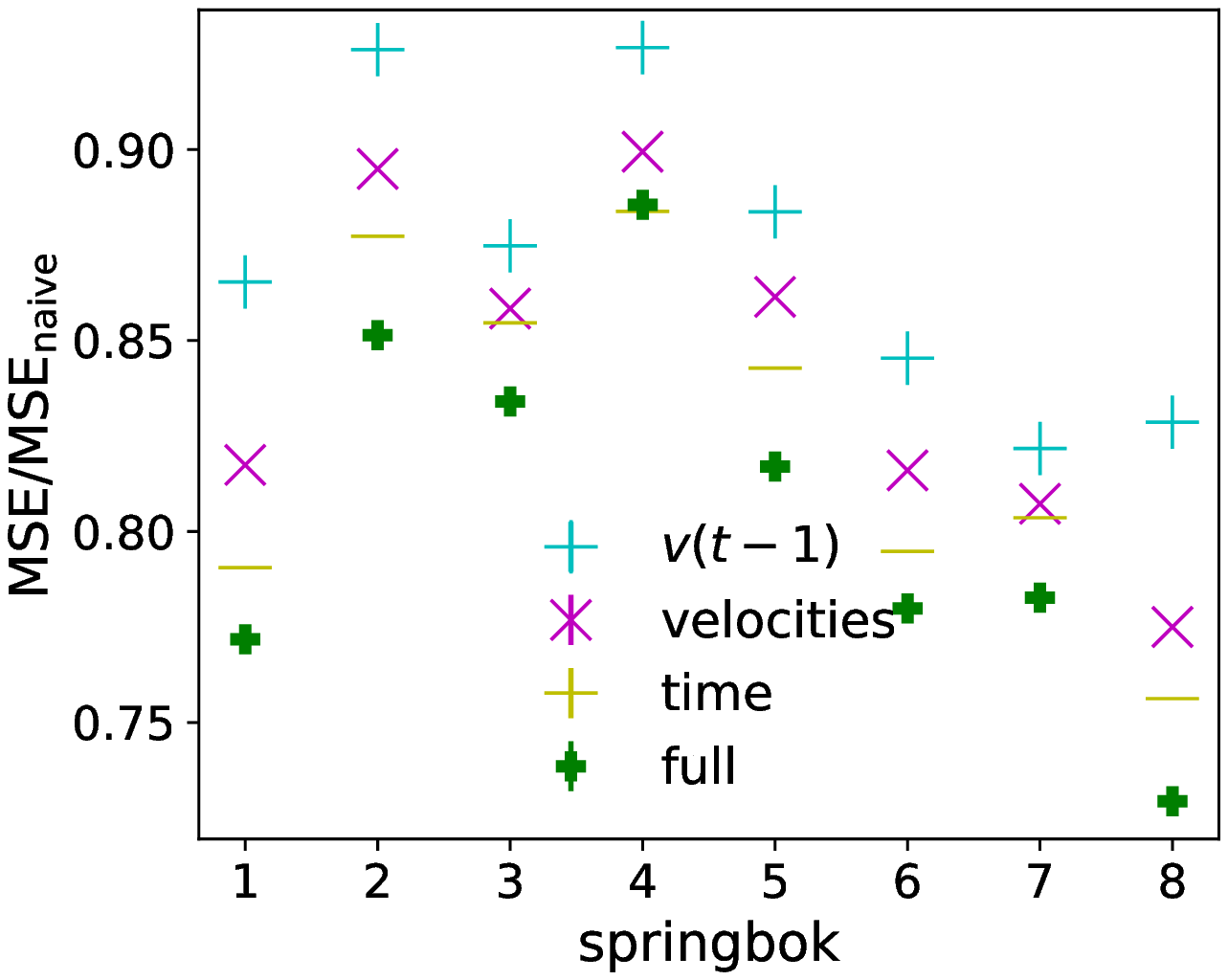}
c)\includegraphics[width=0.31\textwidth]{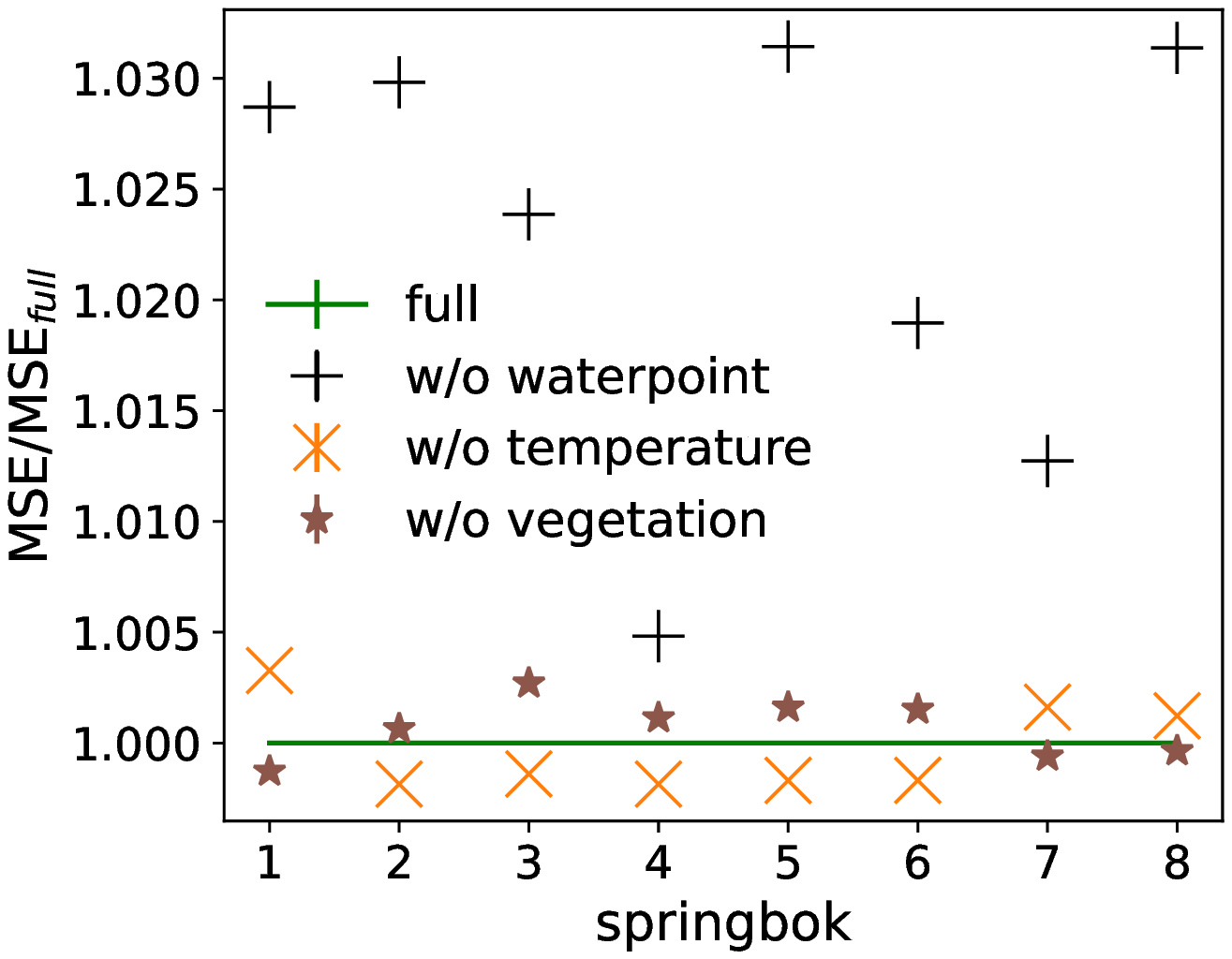}
\caption{MSE of different forecasting models for individual springboks, which
are distinguished by the numbers on the axis. (a) Comparison of the basic
model, the stochastic AR(1) model, and the GBT-based model using all features.
(b) Comparison of the mean squared error, normalized by the prediction of
the basic model for GBT-models with information of the last step $v(t-1)$,
movements on the last day (velocities), movements on the last day and time
of the year (time), and all available information (full). (c) Comparison
of the error normalized by the error of the full model versus the models
where one of the features was removed. Note that in cases when no vertical
lines are visible, the corresponding error is negligibly small.}
\label{fig_pred}
\end{figure*}

\begin{table}
\begin{tabular}{|c|c|c|c|}
\hline
Model & MSE $[\mathrm{km}^2/\mathrm{h}^2]$ &
MSE/MSE$_\mathrm{basic}$ & MSE/MSE$_\mathrm{full}$ \\ \hline
basic & 0.264 & 1.0 & 1.257 \\
AR(1) & 0.228 & 0.873 & 1.087 \\
AR(2) & 0.228 & 0.873 & 1.087 \\
$v_{t-1}$ & 0.228 & 0.872 & 1.082 \\
velocities & 0.219 & 0.841 & 1.044 \\
time & 0.215 & 0.825 & 1.023 \\
full & 0.210 & 0.795 & 1.0 \\\hline
\end{tabular}
\caption{Mean squared error and relative error for different models of springbok
movement. The values are averages over all individuals in the dataset. As seen
from the third column the full model established here reduces the error in
springbok move prediction by around 20\%, as compared to the basic model.}
\label{tab_pred}
\end{table}

As we are interested in forecasting the movement to be made by a springbok in
the next hour ($\Delta t=1$ h), encoded by the "velocity" (distance per unit
time of 1 h)
\begin{equation}
\mathbf{v}(t_i)=\frac{\mathbf{r}(t_i)-\mathbf{r}(t_{i-1})}{\Delta t},
\end{equation} 
the simplest prediction is to assume that, most probably, the next
position is equal the current position. Then the predicted velocity
$\tilde{\mathbf{v}}(t_i)$ vanishes, i.e., $\tilde{\mathbf{v}}(t_i)=0$. Viewing
$\mathbf{v}$ as an averaged quantity, this simplified assumption corresponds
to the picture of Brownian motion, in which steps in either direction are
equally probable, and, as already noted by Pearson \cite{pearson1}, the next
displacement is zero. How realistic this simple prediction or similar ones
are can be quantified in terms of the mean squared error (MSE), evaluated
as squared deviation of the forecast $\tilde{\mathbf{v}}$ with respect to
the actual experimental time series $\mathbf{v}$,
\begin{equation}
\mathrm{MSE}=\frac{1}{L-1}\sum_{i=1}^{L-1}\left(\mathbf{v}(t_i)-
\tilde{\mathbf{v}}(t_i)\right)^2.
\end{equation}
MSEs, or normalized MSEs when compared to a given model, are common measures
for the quality of models in modeling analyses, see, e.g., \cite{henrik,
munoz2021objective}.

A better prediction is expected if the information contained in model
(\ref{eq_model}) is taken into account. For short lag times of around one
hour, confinement effects can be neglected, as these enter only with a
comparatively long correlation time $\tau_l$, which is on the order of tens
of hours, according to Fig.~\ref{fig_tsd}. The resulting model is then a
random walk with correlated increments, characterized by the correlation
time $\tau_s$. Thus, at such short lag times the prediction is given by
\begin{eqnarray}
\label{modcorr}
\mathbf{v}(t_i)=e^{-\Delta t/\tau_s}\mathbf{v}(t_{i-1}),
\end{eqnarray}
in which each component corresponds to an autoregressive model of order one,
AR(1). It can be tested that the full AR(2) model (\ref{eq_model}) does not
lead to any significant improvement of the predictions (see table
\ref{tab_pred}). In fact, model (\ref{modcorr}) already represents a
significant improvement as compared to the naive Brownian prediction and
explains around 13\% of the squared error (see table \ref{tab_pred}).

\subsection{Machine-learning approach}

However, as mentioned above, model (\ref{modcorr}) is still quite simplistic
and neither takes into account the specific animal movements at certain times
of the day or year nor the position of the water points, the dependence on
explorations during the previous day, as well as other factors (such as
variable levels of vegetation or fluctuating temperatures). Obviously, a
general model that could take such features into account is hard to formulate
explicitly. Therefore it is useful to look for suitable machine learning
algorithms. These methods \cite{karn19rnn-dl} are superior to many classical
techniques for model selection for the datasets generated from stochastic
processes \cite{thapa,fabr18, munoz2021objective,aliv21dl-pnas,henrik}. They
can also be used to infer the dynamical changes and the transition points
\cite{song2022machine}. A direct forecast of future steps of an animal
using ML is less common in studying diffusion and foraging (for a review
of potential methods we refer the reader to Ref. \cite{Dormann18}), but it
is already wide-spread in other disciplines such as power-grid frequency
\cite{KRUSE2021100365} or air-pollution research  \cite{cabaneros2019review}.
In these disciplines, the question "how important is the individual feature
for the accuracy of modeling" is often being discussed. The same techniques
can be applied for time-series analysis of animal movement. We deploy a
supervised-ML model, which takes the following features into account:
\begin{itemize}
\item[(i)] $\mathbf{v}(t_{i-1})$, the distance per time covered by an animal in
the last hour (i.e., for a lag time $\Delta t=1$ h)
\item[(ii)] $\mathbf{v}(t_{i-2})$, the distance per time covered in the hour
before that
\item[(iii)] $\mathbf{v}(t_{i-24})$, the distance per time walked in the same
hour of the previous day
\item[(iv)] $\mathbf{r}(t_{i-1})-\mathbf{r}(t_{i-24})$, the overall distance
directions traveled in the last 24 hours
\item[(v)] $w(t_{i-1})$, the distance to the closest water point, and $w(t_{
i-1})-w(t_{i-2})$, the change in distance from the closest water point during
the last hour
\item[(vi)] $T(\mathbf{r}(t_{i-1}))$, the temperature at the current position
\item[(vii)] $V(\mathbf{r}(t_{i-1}))$, the vegetation type as a "lushness
coefficient" (varying in the interval $[0,1]$) at the current position, and
the lushness difference $V(\mathbf{r}(t_{i-1}))-V(\mathbf{r}(t_{i-2}))$ as
compared to the previous location.
\end{itemize}

\begin{figure}
\includegraphics[width=0.45\textwidth]{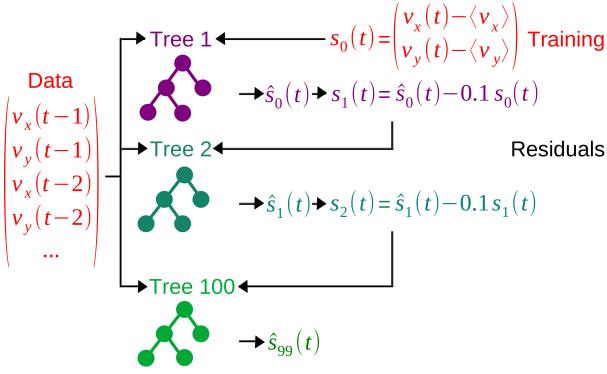}
\caption{Schematic diagram of the GBT-algorithm: the first tree receives as
input the residuals $s_0$ of the inputs and the average result of the training
set. Depending on the specific input values, the tree's output, the estimator
$\hat{s}_0$ can take different values (one of the branches of the tree), which
are trained from the results in the training set $(v_x(t),v_y(t))$, that ended
up on the same branch of the tree. The output is regularized by the learning
rate $0.1$. The residuals $s_1$ are passed as an input to the next tree, which
works in the same way. The maximum number of trees in our application was 100.
After the model is trained the output can be calculated from a new input as
$(\hat{v}_x(t),\hat{v}_y(t))=(\langle v_x\rangle,\langle v_y\rangle)+0.1(\hat{
s}_0(t)+\hat{s}_1(t)+\ldots+\hat{s}_{99}(t))$, where $(\langle v_x\rangle,
\langle v_y\rangle)$ are the mean displacements in the training set.}
\label{fig_gbtdraw}
\end{figure}

Gradient-boosted trees (GBTs) represent a popular tool for regression
analyses \cite{GBTfriedman,GBTnatekin}. GBTs are based on simple decision
trees, and the output of one tree is then passed to the next decision
tree, which learns how the residuals can be improved with additional
questions. The result of each tree is regularized by a factor, called the
learning rate. Choosing a sufficiently small learning rate reduces the risk
of over-fitting \cite{KRUSE2021100365}. We use a learning rate of $0.1$,
which is a standard choice for gradient-boosting machines. The number of
trees and the maximal depth of the trees are the parameters to be chosen
prior to starting the algorithm. We use $n=100$ trees with a maximal depth
of 5, which was decided after comparing the accuracy of the algorithm for
different sets of parameters. We employ the Python scikit-learn library
for a supervised ML-model implemented as GBTs. The black-box
character of such a model could be understood via  running the model with
different features acting as inputs and comparing the importance of each
feature, see below. We note that in this GBT approach we do not use any
specific information regarding the physical position of the springboks, but
only account for their relative distances. The algorithm is thus applicable
to all springboks independently of whether or not the area has already been
explored previously. The data on vegetation levels and the data on temperature
variations are taken from \cite{heringdon}.

In Fig.~\ref{fig_pred} we compare the MSE of the basic (Brownian) model, the
prediction from the AR(1) model (\ref{modcorr}), and the supervised ML-model.
For the latter two, the models were trained on all but one individual and
the next step was predicted for this remaining individual. In the case of the
stochastic model, only the parameter $\tau_s$ had to be "learned": it can be
obtained from minimizing $[\mathbf{v}(t_i)-e^{-(1h)/\tau_s}\mathbf{v}(t_{i-1}
)]^2$. The stochastic model reduces the MSE of the prediction by $\approx$13\%
on average. Using all available features in the GBT model leads to a
total reduction the MSE of $\approx$20\%, see the entry 0.795 in the third
column of Tab.~\ref{tab_pred}. Consequently, the
springbok motion contains a significant degree of stochasticity that cannot
be explained even with the quite broad features of the GBT approach. We
also mention the measurable variability of the MSE-results computed for
movements of different individuals in Fig.~\ref{fig_pred}, resulting from
characteristics like age, size, gender, distinct set of explored terrain
patterns, movement strategy (residential vs. migrating), etc..

\subsection{Feature gain}

ML-models such as GBTs can be considered as "black boxes": the effects of an
individual feature are not immediately obvious. The easiest way to gain some
insight into such a black box is to run the algorithm repeatedly and to
examine the effects of adding or deleting certain individual features.
More information should lead to a better prediction of the future
behavior, but the question is how relevant an individual feature is for such
predictions.

In Fig.~\ref{fig_pred}b we demonstrate how subsequently adding information
reduces the MSE. Errors are normalized by the error of the basic model (the
variance of the velocity). It can be shown that taking into account only the
previous data point---i.e., the same amount of information considered for the
stochastic forecast---very well reproduces the MSE of the stochastic model,
$\approx$13\% (see Table \ref{tab_pred}). Adding information about the
springbok paths during the previous hour and on the previous day, $\mathbf{v}
(t_{i-2})$, $\mathbf{v}(t_{i-24})$, $\mathbf{r}(t_{i-1})-\mathbf{r}(t_{i-24})$
as three additional features yields a strong improvement of the error: it is
almost 16\% better than the basic model. Also, taking into account the time of
the day and and of the year gives rise to another visible improvement, as can
be expected from the changes of the dynamics depending on time discussed above.
The information on changes in the distance to the water points has some effect,
while the vegetation
level, and the actual temperature lead to hardly any additional improvement.
One reason might be, that some of this information is already 
included in the movement on the previous day, e.g. if the animal already 
visited the same water point or went to the same pasture to eat.

Looking at these three factors (previous data point, longer movement history,
time of day/year) in detail, we can also consider the relative increase of
the MSE upon removal of either features from the full model, as shown in
Fig.~\ref{fig_pred}c. We see that there is only one of the three factors that
has a negative effect throughout all individuals, namely, the distance from
the water points. The magnitude of the effect is, however, rather small. It is
not clear that the model benefits much from information on the vegetation and
temperature. For the vegetation, this is not an expected result, however, in
order to predict where the animal goes to eat one would need more information
about the surroundings and a higher temporal and spacial resolution.

\section{Conclusion}
\label{conc}

We studied the movement dynamics of an ensemble of springboks, whose positions
were recorded by long-term GPS tracking. Our analysis combines new statistical
observables with stochastic models and ML-based feature analysis. Although the
studied springbok ensemble is relatively small and the statistical quality of
our results is therefore limited, we believe that this study will be a solid
basis for more elaborate experimental field work and extraction of dynamic
features from the garnered data. In particular, all techniques discussed here
should be easily adaptable to the analysis of the movement dynamics of other
ruminants. With sufficient modifications, possibly including the underlying
stochastic models, the movement dynamics of other tracked animal species
can be approached with the same methodology. In this sense our analysis presents
a next step from the purely statistical description of animal movements, in the
direction of a more detailed, biologically-inspired analysis and prediction.

The evaluation of the model performance ("forecasting error") for the movement
in the following hour was based on the MSE obtained from comparison of the
predicted movement of a given animal based on the model (after training of the
model parameters from all other individuals) with actual animal movement.
A simple model chosen for the movement dynamics was the discrete OU-process with
correlated driving, corresponding to the autoregressive model AR(2). This process
includes the confinement of the animal motion over time ranges of some $10^2$ h
corresponding to a few km, as seen from the detailed TAMSD data. For the movement
within the chosen 1 h lag time for the prediction analysis, the AR(2) model
reduces to the AR(1) process, an unconfined, correlated motion. This model was
shown to already have a quite good forecasting power, leading to a reduction of
the MSE as compared to a basic model, in which animals on average stay put at
their current positions. The gain in prediction from the correlated motion model
was $\approx13\%$. The error did not vary appreciably between this AR(1) model,
the full AR(2) model, and even the GBT model, when the latter was solely trained
on the previous position.

Naturally, a homogeneous model such as AR(1) with two parameters (diffusivity
or noise strength and correlation time $\tau_s$) misses many relevant features
of the real animal movement dynamics. A prime feature here is the daily cycle
in the  animal behavior. Using a decomposition technique based on the scaling
exponents of total displacement, activity, and directedness, shows that not only
the activity, but also the directedness of the springbok motion is higher during
the day (the corresponding exponent is $\approx1.5$) than during the night
(with exponent $\approx1.3$).

Taking into account the step at the same hour on the previous day, the daily
displacement, and the last two steps, the error of the prediction is reduced
by $\approx16\%$ as compared to the basic model, as some of the daily dynamics
is captured. The prediction is even better when information about the time of
the year is added as well, because there is a difference between the behavior
during wet versus dry seasons. A model with information about the time of
day and of the year improves the naive prediction with the basic model by
$\approx18\%$. Note
that such models do not have any specific information about the underlying
map and are thus purely dynamical formulations. An important feature of
the concrete physical landscape were shown to be the water points. While
the scatter of the visited water points is higher than that of the resting
points, the information of the water point distances turned out to have
a clear effect. In contrast, including the current temperature did not
improve the model prediction, similar to the lushness and its gradient.
However, it is not immediately clear to which extent the information about
temperature bears on the movement dynamics at all. Moreover, it would be
reasonable to assume that the geographic abundance of food is encoded in
the movements themselves.  For an improved model, one could consider the
food landscape in the entire field of vision of the animals and their memory.

Our best ML-based model with all information improved the MSE by about 20\%.
While this points at the relevance of the underlying parameters it also
shows the limitation of the predictability, pointing at a good degree of
stochasticity of individual motion. We propose that the reasons for this are
not only the heterogeneity among the individuals reflected in variability
of their movement parameters, but also in the multiple decisions taken in
dependence of the current needs of an individual and its interactions with
other animals, as well as personal perceptions. Averaging over different
individuals from different herds leads to a loss of details of the individual
animal movement.  Concurrently, such averaging unveils features such as the
motion directedness during day and night, scattering of water points and
resting positions, and the predictability of the springbok movement without
any knowledge of the underlying landscape. When more detailed data will be
available, we will be able to extend our analysis on exact geographical
features such as detailed vegetation and height maps, as well as the
positions of fences and of other animals in the herd. In such an analysis
also the time resolution of water availability at water points can be taken
into consideration. However, it is an open questions whether such details
are actually that important, or whether it is sufficient to have knowledge
about more generic motion patterns.

We conclude by speculating that a substantial improvement of the relatively
simple, few-parameter approaches outlined here, in combination with machine
learning, may still be achievable. Namely, from feature-based machine learning
studies we know that few additional features, on top of a substantial number
already used, may significantly improve the predictions. Here AI combining all
available information may be used in the future to pinpoint additional relevant
factors in movement ecology.

\appendix

\section{A primer on the Mandelbrot decomposition method}
\label{mosesapp}

The central limit theorem for the sum of random variables guarantees the
convergence to a Gaussian PDF if these random variables are independent,
identically distributed, and of finite variance \cite{hughes}. Violation
of any of these conditions effects changes to the resulting PDF. When a
diffusive process deviates from the Gaussian statistic of normal Brownian
motion, a direct test of which of these conditions is violated, can be
based on three scaling exponents defined on the basis of the increments of
the position time series \cite{mand68moses,che17,meyer18,aghion2021moses,vilk}.

Consider the position time-series $\mathbf{r}(t)$ as function of time $t$,
described by the discrete sum of random increments $\mathbf{r}(N\Delta t)=
\sum_{j=1}^N\delta \mathbf{r}_j$, where $\delta \mathbf{r}_j\equiv\mathbf{r}
(j\Delta t)-\mathbf{r}([j-1]\Delta t)$ and $N=t/\Delta t$, and $0<\Delta t\ll
t$ is an arbitrary time increment. We also define the average velocity vector
in the $j$th increment, $\mathbf{v}_j\equiv\delta\mathbf{r}_j/\Delta t$. From
these quantities we calculate the scaling of three observables: (i) the mean
absolute velocity $\langle|\mathbf{v}|\rangle$, (ii) the mean squared
velocity $\langle\mathbf{v}^2\rangle$, and (iii) the ensemble-averaged
time-averaged MSD (TAMSD). These are connected to the following effects:

(i) Non-stationarity. The conditions that the random variables are
identically distributed is violated. Non-stationarity of the increments can
be measured by the the "Moses" exponent $M$ defined in terms of
\begin{equation}
\left<\overline{|\mathbf{v}|(N\Delta t)}\right>=\left<\frac{1}{N\Delta t-
\Delta t}\sum_{j=1}^{N}|\mathbf{v}_j|\right>\propto N^{M-1/2},
\label{eq1}
\end{equation}
where the overline indicates time averaging (TA). When $M=1/2$ the increments
are stationary and thus identically distributed. When $M>1/2$ the process is

(ii) Diverging variance. Extreme events in the time series are picked up by
the "Noah" exponent $L$ (generally, $L\geq 1/2$ \cite{mand68moses}), given
through
\begin{equation}
\left<\overline{\mathbf{v}^2(N\Delta t)}\right>\equiv\left<\frac{1}{N\Delta t
-\Delta t}\sum_{j=1}^{N}\mathbf{v}^2_j\right>\propto N^{2L+2M-2}.
\label{eq2}
\end{equation}
If in addition to $M=1/2$ we have $L=1/2$, then $\langle\mathbf{v}^2\rangle$
is constant. Yet, if $L>1/2$, its value will grow in time, even though $M=1/2$.
While $\langle|\mathbf{v}|\rangle$ is a measure for the typical fluctuations in
a time series, the quantity $\langle\mathbf{v}^2\rangle$ is sensitive to the
tails of the velocity PDF. In absence of extreme events $\langle\mathbf{v}^2
\rangle\propto N^{2M-1}$, thus $L\neq1/2$ is an indicator for the occurrence
of extreme events.

(iii) Temporal correlations. The "Joseph" exponent measures whether the
increments of the process are independent or not. This exponent can be
extracted from the scaling of the integrated velocity autocorrelation
function
\begin{equation}
\frac{1}{\langle\mathbf{v}^2\rangle}\sum_{i=1}^l\langle\mathbf{v}_i\cdot
\mathbf{v}_{i+l}\rangle\propto l^{2J-1}.
\end{equation}
As this function is sometimes difficult to calculate from finite time series,
an alternative definition is based on the mean time-averaged MSD
\begin{equation}
\left<\frac{1}{N-l}\sum_{i=1}^{N-l}\left(\mathbf{r}(t_{i+l})
-\mathbf{r}(t_i)\right)^2\right>\propto N^{2L+2M-2}l^{2J}.
\end{equation}
For long-ranged temporal correlations (decaying very slowly in time) one has
$J\neq1/2$, thus violating the independence condition of the central limit
theorem.

There exists a fundamental summation relation between $M$, $L$, $J$ and the
Hurst exponent $H$ \cite{che17,meyer18,aghion2021moses},
\begin{equation}
\label{summation_relation}
H=J+L+M-1.  
\end{equation}
This relation was empirically confirmed in a wide range of systems \cite{vilk}.
An observed process resembles Brownian motion when $L=M=J=H=1/2$. As we do not
observe a growing variance $\langle\mathbf{v}^2\rangle$, and thus we assume $L
=1/2$ in the main text.

The summation relation (\ref{summation_relation}) for $L =1/2$ can be derived as
follows. Assuming a power-law scaling dependence of the autocorrelation function
with the lag time $\Delta$ and the measurement time $t$, we have (for the
$x$-component)
\begin{eqnarray}
\nonumber
\langle(x(t)-x(0))^2\rangle&=&2\int_0^td\Delta\int_0^{t-\Delta}dt'\langle v(t')
v(t'+\Delta)\rangle\\
\nonumber
&\propto&2\int_0^t d\Delta\int_0^{t-\Delta}dt'\langle v^2(t')\rangle\Delta^{2J-2}\\
\nonumber
&\propto&2\int_0^td\Delta\Delta^{2J-2}\int_0^{t-\Delta}dt'{t'}^{2M-1}\\
&\propto&\frac{B(2J-1,2M+1)}{M}t^{2M+2J-1},
\label{eq_ConnectionBetweenExponents1}
\end{eqnarray}
where the Euler Beta-function is given by
\begin{equation}
B(\alpha,\beta)=\int_0^1t^{\alpha-1}(1-t)^{\beta-1}dt.
\end{equation}
The last step requires, that $J>1/2$ and $M>0$, which is true in our
situation. In this case, relation (\ref{eq_ConnectionBetweenExponents1})
holds for all $t$ and $\Delta$, for smaller exponents it only holds in
the long-time limit.

\begin{acknowledgments}

We acknowledge funding from the German Science Foundation (DFG, grant number
ME 1535/12-1). We also acknowledge the Research Focus "Data-centric sciences"
of University of Potsdam for funding. 
Springbok movement data acquisition was  funded in the ORYCS project
within the SPACES II program, supported by the German Federal Ministry
of Education and Research (grant number FKZ 01LL1804A).

\end{acknowledgments}

\section*{Competing interests}

The authors have no competing interests to declare.

\section*{Data availability}

The data-set is available upon reasonable request from the authors.

\end{document}